\documentclass{aastex701}

\usepackage{amsmath}
\usepackage{booktabs}

\begin{document}

\title{Long-Term Dynamical Evolution and Ejection of Near-Earth Asteroids}

\author[orcid=0009-0008-9957-0375]{Chetan Abhijnanam Bora}
\affiliation{Department of Mathematics and Computing, Indian Institute of Technology (ISM) Dhanbad, Dhanbad, Jharkhand 826004, India}
\email[show]{chtnbora@gmail.com}
\correspondingauthor{Chetan Abhijnanam Bora}

\author[orcid=0000-0001-9569-2297]{Badam Singh Kushvah}
\affiliation{Department of Mathematics and Computing, Indian Institute of Technology (ISM) Dhanbad, Dhanbad, Jharkhand 826004, India}
\email{bskush@gmail.com}

\author[orcid=0000-0002-8768-9298]{Kanak Saha}
\affiliation{Inter-University Centre for Astronomy and Astrophysics (IUCAA), Pune 411007, India}
\email{kanak@iucaa.in}

\begin{abstract}
Long-term integrations of asteroid orbits with high-accuracy numerical integrators are essential for understanding dynamical evolution and ejection from the Solar System, but are computationally expensive. Here, we investigate the dynamical behaviour of asteroids and explore machine-learning (ML) and deep-learning (DL) approaches as efficient, scalable alternatives for classifying long-term dynamical outcomes. While the ML classifiers are trained on initial orbital elements, the convolutional neural network is trained on recurrence plots derived from short-period numerical integrations generated with the MERCURY integrator. Ensemble tree models perform strongly on the ephemeris input, and the neural network captures temporal signatures of chaotic motion with comparable or slightly improved accuracy. Backward integrations reveal partial overlap between forward- and reverse-ejected sets, illustrating time-asymmetric behaviour in chaotic regions; these backward results are interpreted only as diagnostic probes rather than reconstructions of past histories. Non-ejected asteroids largely correspond to known dynamical groups, underscoring the constraining role of initial orbital configuration. These methods provide scalable frameworks to complement numerical integrations and inform prioritisation for detailed long-term dynamical studies, with implications for planetary-defence analyses.
\end{abstract}

\keywords{\uat{minor planets, asteroids: general}{} --- \uat{astronomical databases: miscellaneous}{} --- \uat{methods: data analysis}{} --- \uat{methods: numerical}{} --- \uat{methods: statistical}{} --- \uat{celestial mechanics}{}}


\section{Introduction}

Asteroids are primordial remnants from the early Solar System that offer valuable insights into its formation and evolutionary history. While most reside in the main asteroid belt between Mars and Jupiter, a significant fraction occupy more perturbed orbits that bring them into the vicinity of Earth \citep{tabachnik2000asteroids}. These Near-Earth Asteroids (NEAs) \citep{bottke2000understanding}, particularly those with high eccentricities and inclinations, can cross planetary orbits and in some cases pose a threat of impact \citep{bottke1993collision, carruba2025invisible}. The study of their long-term orbital evolution is central to understanding not only their origin \citep{morbidelli1999origin} and fate but also their potential to become impact hazards \citep{chapman2004hazard, perna2013near}. One particularly critical question concerns whether such objects remain gravitationally bound to the Solar System or are eventually ejected \citep{levison1994long, gladman2000near, granvik2018debiased, zolotarev2021dynamic}.

High-precision $N$-body integrations are commonly used to investigate the long-term gravitational evolution of small bodies over Myr--Gyr timescales \citep{everhart1985efficient, duncan1998multiple, chambers1999hybrid, rein2012rebound}. A variety of numerical schemes are employed in such studies, broadly including symplectic and high-accuracy non-symplectic integrators. Symplectic methods conserve a modified Hamiltonian and therefore exhibit favourable long-term energy behaviour \citep{yoshida1990construction}. High-accuracy non-symplectic integrators (e.g., Bulirsch--Stoer, RADAU, IAS15) employ adaptive timestepping with strict local-error control, enabling very accurate integration of the equations of motion across a wide range of dynamical regimes \citep{bulirsch1966numerical, rein2015ias15}. Such integrations remain computationally intensive, particularly when applied to large asteroid populations. The resulting datasets, while costly to generate, provide a foundation for recent data-driven approaches that seek to build predictive models using ML and DL. The dynamical evolution of NEAs is strongly chaotic, with typical Lyapunov times $\tau_{L}$ of order $10^{2}$\,yr (commonly $\sim$100--200\,yr), much shorter than the 1\,Myr analysis window \citep{wisdom1987urey,milani1997stable}. Consequently, we interpret all integrations and model outputs in this study in a statistical, ensemble-level sense rather than as deterministic orbital reconstructions.

ML techniques have been widely adopted in asteroid research, particularly where structured input features such as orbital elements are available \citep{baron2019machine}. They have been successfully applied to classify asteroid families using proper elements \citep{carruba2019machine, carruba2020machine, vujicic2020classification}, identify asteroids in resonance \citep{smirnov2017identification, smirnov2024comparative, smirnov2026implementation}, and categorize NEAs into dynamical subgroups such as Apollo, Amor, Aten, and Atira (also known as Apohele) \citep{mako2005classification}. ML classifiers have also been used to assess impact risk by predicting whether asteroids are potentially hazardous, using features such as Minimum Orbit Intersection Distance and absolute magnitude \citep{bahel2021supervised, sharma2023asteroid, malakouti2023machine, bora2024temporal}.

Complementing these efforts, deep learning architectures have been employed in applications involving more complex, high-dimensional, or unstructured data \citep{lecun2015deep, goodfellow2016deep}. Artificial neural networks (ANNs) and Convolutional Neural Networks (CNNs) have been used for dynamical classification tasks in resonant populations \citep{carruba2021artificial, carruba2023deep, carruba2024digitally, carruba2025vision, carita2024image, alves2025deep}, while probabilistic neural networks (PNNs) have predicted orbital stability using integrated trajectory data \citep{liu2021stability}. In the observational domain, CNNs have proven effective in detecting asteroids from astronomical images, with \citet{rosu2021asteroid} achieving a 94\% recall rate, and \citet{bacu2023assessment} showing superior performance of InceptionV3 among several deep CNN architectures. More recently, transformer-based models have been introduced for astronomical time-series classification, with vision transformers applied to recurrence plot (RP) representations of Kepler light curves achieving high-precision exoplanet detection \citep{choudhary2025exoplanet}.

Despite growing interest in data-driven modeling, the application of ML and DL techniques to predict asteroid ejection remains limited. To address this, we propose a dual-framework that independently applies both traditional ML algorithms and CNN-based models to assess the ejection likelihood of NEAs. The ML component utilizes initial orbital elements such as semi-major axis, eccentricity, and inclination as input features to capture the role of initial conditions. We evaluate a diverse set of classifiers, including Decision Trees \citep{quinlan1986induction}, Random Forests (RFs) \citep{breiman2001random}, K-Nearest Neighbors (KNNs) \citep{peterson2009k}, Extremely Randomized Trees (ExtraTrees) \citep{geurts2006extremely}, Multilayer Perceptrons (MLPs) \citep{gardner1998artificial}, Adaptive Boosting (AdaBoost) \citep{zhou2012ensemble}, and Gradient Boosting (GB) \citep{swamynathan2017mastering, schapire2013boosting}.

In parallel, our CNN-based approach operates on RPs generated from short segments of numerically integrated trajectories. These image-based encodings represent dynamical complexity and allow CNNs \citep{lecun2002gradient, krizhevsky2012imagenet} to extract features indicative of chaotic evolution and instability. By integrating both static (initial conditions) and dynamic (trajectory-derived) features, the proposed methodology provides a comprehensive and scalable alternative to exhaustive long-term numerical simulations.

The remainder of this paper is organised as follows. Section~\ref{sec:challenges_ml} discusses the challenges in predicting long-term NEA dynamics and motivates the proposed approach. Section~\ref{sec:data_preprocessing} details the dataset construction, numerical integrations, and preprocessing steps. Section~\ref{sec:recurrence_plots} introduces the recurrence-plot representation of orbital evolution. Section~\ref{sec:methods} presents the machine-learning and CNN frameworks. Section~\ref{sec:classification_results} reports the classification results and dynamical analysis. Finally, Section~\ref{sec:discussion_conclusions} presents the discussion and summarises the main conclusions.

\section{Challenges in Long-Term NEA Prediction and ML-Based Solutions}
\label{sec:challenges_ml}

Predicting the long-term dynamical evolution of NEAs is a formidable challenge, particularly when assessing whether objects remain gravitationally bound to the Solar System or undergo ejection due to cumulative perturbations \citep{morbidelli2002modern}. High accuracy N-body integrations accurately resolve these processes but scale poorly to large catalogs \citep{bulirsch1966numerical, levison2000symplectically}. Consequently, efficient data-driven approaches are needed to approximate ejection likelihoods while preserving the essential dynamical information encoded in orbital parameters and short-term evolutionary trajectories.

\begin{figure*}
	\centering
	\hspace*{-1mm} 
	\includegraphics[width=0.8\textwidth]{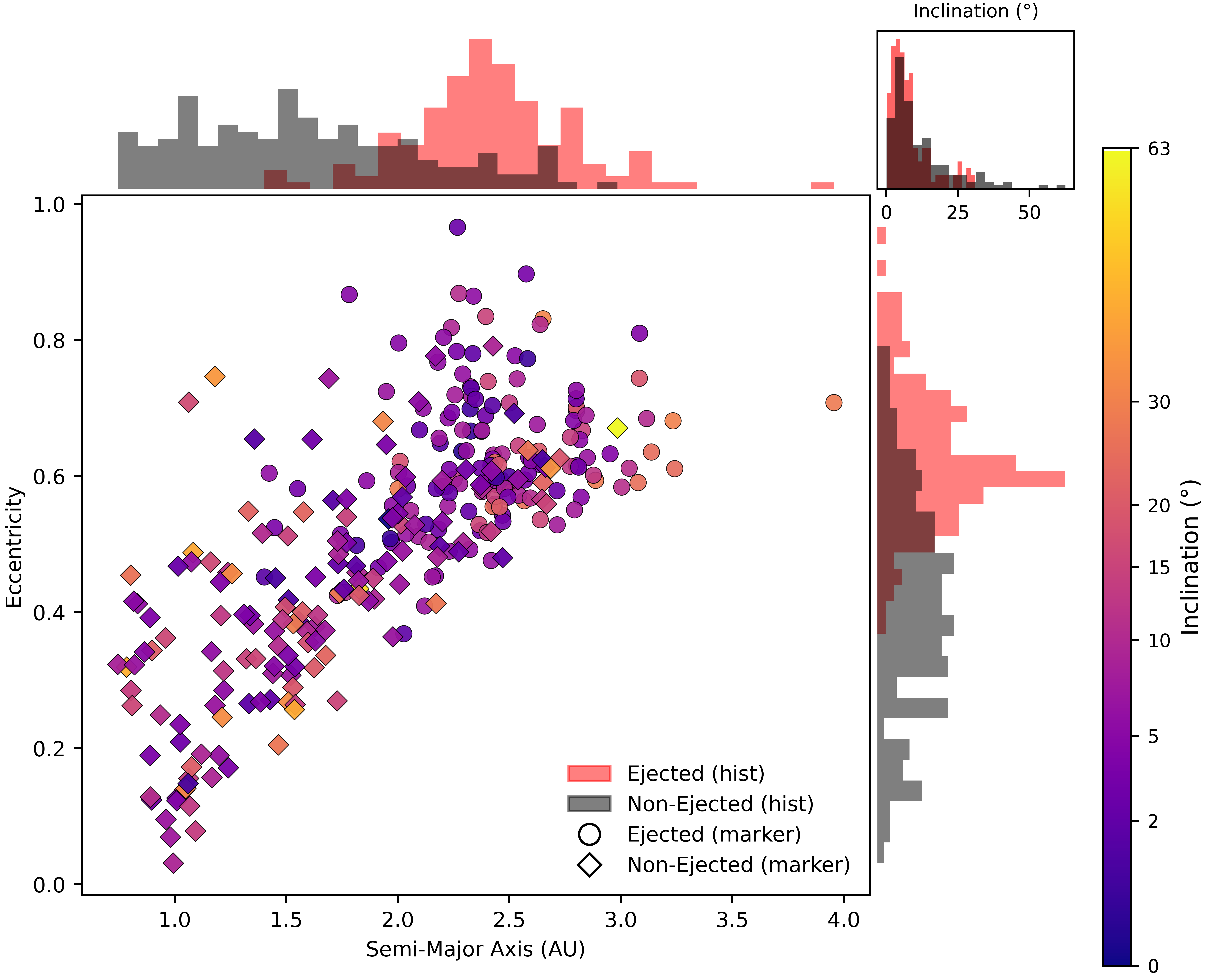}
	\caption{Distribution of ejected and non-ejected asteroids in terms of their semi-major axis and eccentricity, with inclination values represented by the colour scale. The marginal histograms on the top and right illustrate the normalized distributions of the semi-major axis and eccentricity for both classes, while the top-right inset shows the inclination distribution.}
	\label{fig:single_panel_a_e_color_i_with_marginals_longer_hist}
\end{figure*}

Figure~\ref{fig:single_panel_a_e_color_i_with_marginals_longer_hist} illustrates the empirical distribution of ejected and non-ejected asteroids in the space of semi-major axis ($a$) and eccentricity ($e$), with inclination ($i$) represented as a colour gradient. Ejected asteroids cluster predominantly at larger semi-major axes and higher eccentricities regions where orbital instability is theoretically favored due to stronger resonant interactions and enhanced chaotic diffusion \citep{nesvorny2002regular, morbidelli1999numerous}. The marginal histograms confirm this tendency, showing that the ejected population is skewed toward higher semi-major axis and eccentricity values. In contrast, inclination exhibits weaker discriminatory power, with both populations concentrated at low to moderate $i$ ($\lesssim20^{\circ}$). These trends are consistent with classical dynamical theories that link large orbital radii and elongation to increased susceptibility to planetary encounters and long-term instability \citep{gladman1997dynamical}.

Nonetheless, the overlap between ejected and non-ejected objects in the $(a,e)$ plane underscores the intrinsic non-linearity of the ejection problem. Many asteroids occupy similar regions of phase space yet diverge drastically in their eventual dynamical fate. This intermingling indicates that simple linear thresholds or analytical resonance criteria are insufficient to distinguish long-term outcomes, and that non-linear, multi-dimensional relationships among the orbital parameters govern stability. Such complexity strongly motivates the adoption of ML and DL frameworks capable of identifying intricate, non-linear decision boundaries and capturing subtle dynamical correlations invisible to human inspection \citep{carruba2022machine, penttila2021asteroid}.

Within this context, two complementary predictive paradigms are explored. The first leverages conventional ML classifiers such as Decision Trees, RFs, GBs, etc trained on static orbital elements measured at a given epoch \citep{pandey2019machine}. Although these models do not incorporate explicit temporal evolution, they effectively exploit statistical relationships among features like semi-major axis, eccentricity, and inclination to infer dynamical tendencies. These methods serve as computationally efficient baselines, providing fast and interpretable predictions from structured input data readily available in asteroid databases.

The second, and conceptually more powerful, approach employs a CNN trained on RPs derived from short-term orbital integrations. RPs represent the temporal recurrence of states within a dynamical system, transforming one-dimensional time series into two-dimensional spatial patterns that encode periodicity, quasi-periodicity, and signatures of chaotic divergence. CNNs, originally designed for spatial feature extraction \citep{lecun1998convolutional}, are thus ideally suited to learn from these image-based representations of temporal evolution. Their hierarchical convolutional filters capture localized and global patterns such as abrupt dynamical transitions or gradual diffusion in orbital elements that are indicative of long-term stability or ejection. This makes CNNs particularly effective for recognizing nonlinear temporal dependencies and latent structures in complex dynamical systems.

The visual trends observed in Figure~\ref{fig:single_panel_a_e_color_i_with_marginals_longer_hist} provide an intuitive justification for this choice. The distribution of ejected asteroids along gradients of semi-major axis and eccentricity suggests the existence of structured, spatially coherent patterns in the underlying dynamical trajectories, patterns that CNNs can naturally exploit when trained on recurrence images \citep{hsueh2019condition, mathunjwa2022ecg}. In contrast to traditional ML algorithms, which operate purely on scalar inputs, CNNs leverage both spatial and temporal correlations, allowing them to infer how subtle oscillations, resonant interactions, and diffusive behaviours manifest in time-dependent orbital evolution. Such representations are essential for capturing the fine-grained temporal signatures of chaos that often precede ejection events.

In summary, the hybrid framework adopted in this study unites the interpretability and computational efficiency of traditional ML models with the superior representational power of CNNs \citep{janiesch2021machine}. The ML classifiers provide rapid, low-cost screening of asteroid populations based on static orbital descriptors, while the CNN captures the deeper temporal organization inherent in their dynamical evolution. Together, these methods form a scalable, data-driven alternative to exhaustive long-term N-body simulations. The subsequent section details the dataset construction, preprocessing pipeline, and feature generation procedures that underpin both modeling strategies.

\section{Dataset Preprocessing}
\label{sec:data_preprocessing}

To classify long-term ejection outcomes of NEAs, this study employs a hybrid framework that combines classical ML and DL methodologies. The overarching objective is to reduce the computational expense associated with exhaustive long-term N-body integrations by leveraging either initial orbital conditions or short-term dynamical behaviour to forecast whether an asteroid is ejected from the Solar System over a 1~Myr timescale. This timescale is chosen because it spans the characteristic $10^5$--$10^6$~yr dynamical evolution associated with transport through major resonances and repeated planetary encounters, which drive orbital instability and ejection \citep{bottke2002debiased}.

\subsection{N-body Integrations with MERCURY}
\label{sec:mercury}

High-accuracy $N$-body integrations were performed with the MERCURY package \citep{chambers1999hybrid}. Several well-tested $N$-body packages (e.g., SWIFT, REBOUND, GENGA) provide alternative integrators with different trade-offs in long-term behaviour, encounter handling, and parallel/GPU support \citep{levison1994long, rein2012rebound, rein2015ias15, grimm2014genga}. We adopt MERCURY for its flexible integrator suite and built-in support for user-defined forces. All integrations were performed using the high-accuracy Bulirsch--Stoer (BS) integrator within MERCURY (with \texttt{algorithm = bs} specified in the parameter file), which provides adaptive step control and strict local-error tolerance well suited to systems with frequent close encounters. The BS integrator is a non-symplectic extrapolation method that advances each step via the modified midpoint scheme and refines the solution through Richardson extrapolation to the zero-step limit.

The dynamical model includes the Sun, the eight major planets, and Pluto as massive bodies; asteroids are treated as massless test particles. Non-gravitational accelerations are included within the MERCURY framework by applying additional forces along the radial, transverse, and normal directions, thereby capturing Yarkovsky-like thermal forces and related non-gravitational perturbations acting on small bodies. Integration settings are chosen conservatively to ensure long-term numerical accuracy and stability, following the integration protocol adopted in \citet{bora2024temporal}. The integrations employ a primary timestep of $\Delta t = 8$~days, a Bulirsch--Stoer tolerance of $10^{-12}$, and an output cadence of 10~years. Across independent 1~Myr integrations, the relative energy drift at the final epoch remained at the level of $|\Delta E / E| \lesssim 10^{-10}$, consistent with the validation protocol described in \citet{bora2024temporal}. Unless otherwise stated, trajectories were integrated forward for 1 Myr; backward integrations were performed only for diagnostic comparison, and were not used for training. Initial orbital elements are obtained from the NASA JPL Solar System Dynamics database\footnote{\url{https://ssd.jpl.nasa.gov/}}.

Following \citet{chambers1999hybrid}, the Bulirsch--Stoer integrations are carried out in Cartesian coordinates with respect to the central body. The equations of motion for each massless asteroid are written as

\begin{equation}
\dot{\mathbf r} = \mathbf v ,
\end{equation}
\begin{equation}
\dot{\mathbf v} =
- G m_\odot \frac{\mathbf r}{\lVert \mathbf r \rVert^3}
- G \sum_{j=1}^{N_p} m_j
\left[
\frac{\mathbf r-\mathbf r_j}{\lVert \mathbf r-\mathbf r_j \rVert^3}
+
\frac{\mathbf r_j}{\lVert \mathbf r_j \rVert^3}
\right]
+ \mathbf a_{\rm NG},
\end{equation}

where $\mathbf r$ and $\mathbf v$ denote the heliocentric position and velocity of the asteroid, $m_\odot$ is the solar mass, the sum runs over the massive perturbers (the planets and Pluto), and $\mathbf a_{\rm NG}$ represents the additional non-gravitational accelerations.

Physical collisions were disabled in the present integrations: asteroids were treated as massless test particles, and no merging, removal, or fragmentation occurred upon close approach in the MERCURY N-body integrator. This assumption focuses the analysis on gravitational scattering and resonant dynamics rather than collisional processes.

Given the strongly chaotic nature of NEA dynamics and their short Lyapunov times, our goal is not to forecast exact long-term ephemerides but rather to extract persistent dynamical patterns from short-term trajectories. These patterns remain manifest within the 1~Myr window and can be learned by a CNN to associate dynamical behaviour with eventual ejection or long-term stability.

To assess sensitivity to observational uncertainty, we generated Monte Carlo ensembles for a representative subset of the asteroid sample ($N = 50$ clones per object) by sampling the full $6 \times 6$ covariance matrices reported by the JPL Small-Body Database at the covariance epoch. Sampling was performed in the parameter vector $(e, q, t_p, \Omega, \omega, i)$, subject to basic physical constraints ($0 < e < 1$, $q > 0$, $0^\circ \le i \le 180^\circ$). Conversion to osculating elements employed $a = q/(1 - e)$, with the mean anomaly evaluated at the covariance epoch.

The choice of $N = 50$ clones provides adequate sampling of the local uncertainty region of the nominal orbit while allowing us to test the sensitivity of the dynamical evolution to observational uncertainties. Because the objective of the cloning procedure is to assess the robustness of the macroscopic dynamical outcome (ejection) over a $1\,\mathrm{Myr}$ integration, rather than to derive precise statistical probabilities, a moderate number of realizations is sufficient to probe the divergence of nearby trajectories in this chaotic dynamical regime. Tests performed with larger clone ensembles yielded similar fractions of objects undergoing ejection or remaining bound to the Solar System, indicating that the qualitative dynamical conclusions are insensitive to the exact number of clones used. Thus, $N = 50$ represents a practical compromise between adequate sampling of the orbital uncertainty region and computational cost. Each clone ensemble was then integrated using identical numerical settings in the MERCURY integrator. We found that, on average, more than 75\% of the clones associated with a given object shared the same macroscopic dynamical outcome (i.e., ejection or long-term bounded motion), demonstrating that the inferred dynamical classifications are robust to realistic observational uncertainties.

\subsection{Ejection Criteria and Label Definition}
\label{sec:ejection}

An asteroid is classified as \emph{ejected} if, at any point during the 1~Myr integration window, it exits the Solar System. Operationally, this condition can be satisfied if the asteroid exceeds a heliocentric distance threshold of 100~AU or attains hyperbolic escape velocity. Objects that do not meet either of these conditions throughout the integration are labeled as non-ejected.

This binary classification (ejected versus non-ejected) serves as the target label for all subsequent ML and DL models. The use of a 100~AU threshold is consistent with common practice in long-term dynamical studies of small bodies and provides a practical criterion for identifying permanent dynamical removal from the inner Solar System.

Because only a minority of integrated asteroids are ejected during the 1~Myr integration interval, the resulting binary dataset is inherently imbalanced, with substantially fewer ejected objects than long-lived ones. Class imbalance is a well-known challenge in machine-learning applications to asteroid dynamics \citep{carruba2023imbalanced}. In the present work, this imbalance is mitigated by augmenting the minority (ejected) class using the dataset augmentation procedures described in Section~\ref{sec:augmentation}.

\subsection{Time Series Generation and Recurrence Plots}
\label{sec:rp_generation}

To capture signatures of chaotic and non-linear dynamical behaviour, we extract short-term orbital time series from the numerical integrations and transform them into RPs \citep{eckmann1995recurrence, marwan2007recurrence}. For each asteroid, the time evolution of the semi-major axis and eccentricity was sampled at regular intervals over a time window of 0.2~Myr.

Asteroids that are ejected before completing the full time window of 0.2~Myr produce incomplete sequences; these are padded with zeros to preserve uniform input dimensions across the dataset \citep{ismail2019deep}. This approach retains information about early termination of orbital evolution, which is physically meaningful for identifying ejection-prone trajectories within the fixed 0.2~Myr analysis window. Since only a small fraction of objects are ejected within this interval and the majority of inputs contain complete time series, the classifier does not rely on padding patterns but instead learns from the dynamical structure present in the pre-ejection evolution. This strategy yields balanced classification performance while maintaining high recall for ejected objects, which is a primary objective of this study. Any non-numeric entries were converted to NaN values and subsequently replaced with the mean of the corresponding normalized variable.

Prior to RP construction, the semi-major axis and eccentricity time series are normalized independently using min--max scaling to the range $[0,1]$ \citep{patro2015normalization}, according to
\begin{equation}
	x_{i}^{\text{normalized}} = \frac{x_{i} - \min(x)}{\max(x) - \min(x)}.
\end{equation}

Normalization ensures consistency in distance calculations and prevents distortions in the recurrence matrix arising from differences in variable scaling. The normalized semi-major axis and eccentricity sequences are then concatenated into a single column vector, with the semi-major axis preceding the eccentricity, to form a unified representation of the system’s dynamical state. RPs are subsequently constructed from these processed sequences, yielding image-based representations suitable for convolutional neural network input.

\subsection{Dataset Augmentation and Splitting}
\label{sec:augmentation}

Of the 35,455 integrated asteroids, 5,232 experienced ejection during the 1~Myr integration interval. This reflects the class imbalance noted in Section~\ref{sec:ejection}, leaving the ejected population substantially underrepresented. While this dataset is sufficient for training traditional ML classifiers, it is relatively small for DL applications, which typically require larger training samples to achieve stable and generalizable performance \citep{chen2014big}.

To mitigate this limitation, we augment the dataset by generating synthetic NEAs through controlled perturbations of the initial orbital elements of known objects. Small numerical variations are applied to the semi-major axis (typically $\lesssim \pm 0.05$~AU), eccentricity ($\lesssim \pm 0.02$), and inclination ($\lesssim 3^{\circ}$), ensuring that the perturbed orbits remain within the formal NEA classification boundaries. These values are chosen to be small relative to characteristic orbital scales, such that the resulting orbits remain physically plausible and sample the local phase-space neighbourhood, while being sufficiently large to expose the learning algorithms to representative dynamical variability. Each perturbed asteroid is subsequently re-integrated for 1~Myr using the MERCURY N-body integrator, and its dynamical outcome is re-labeled based on the resulting evolution. The resulting augmented population preserves the statistical and dynamical properties of the original NEA sample and does not exhibit spurious distribution shifts, thereby providing a physically consistent extension of the underlying NEA phase-space distribution.

Initially we extract 0.2~Myr segments from each 1~Myr integration as the base input for RP construction, which are subsequently used as inputs to the CNN. To increase the number of physically representative training examples, particularly for the minority ejected class, we then partition each full integration into non-overlapping 0.2~Myr windows and treat each window as an independent sample, using pre-ejection windows when an object is removed early \citep{box2015time}. Only asteroids classified as NEAs at the beginning of each interval are retained for inclusion in the augmented dataset. Because the trajectories are produced with a high-accuracy $N$-body integrator \citep{chambers1999hybrid}, which preserves the qualitative phase-space structure over the integration interval, temporally distinct windows constitute dynamically consistent short-term realizations suitable for training. Care is taken to avoid overlapping or intersecting windows across training/validation/test splits to prevent data leakage. Temporal segmentation of time series is a standard augmentation strategy in deep learning for time-dependent data and has been used widely to improve CNN generalization by exposing the network to varied dynamical phases \citep{goodfellow2016deep, ismail2019deep}.

This procedure yields a more comprehensive and statistically representative collection of RPs, improving the robustness and applicability of the DL models while maintaining physical consistency in the underlying orbital dynamics.

\section{Recurrence Plots and Feature Representation}
\label{sec:recurrence_plots}

\begin{figure*}
\centering

\begin{minipage}{0.44\textwidth}
    \centering
    \begin{minipage}{0.9\linewidth}
        \centering
        \includegraphics[width=\linewidth]{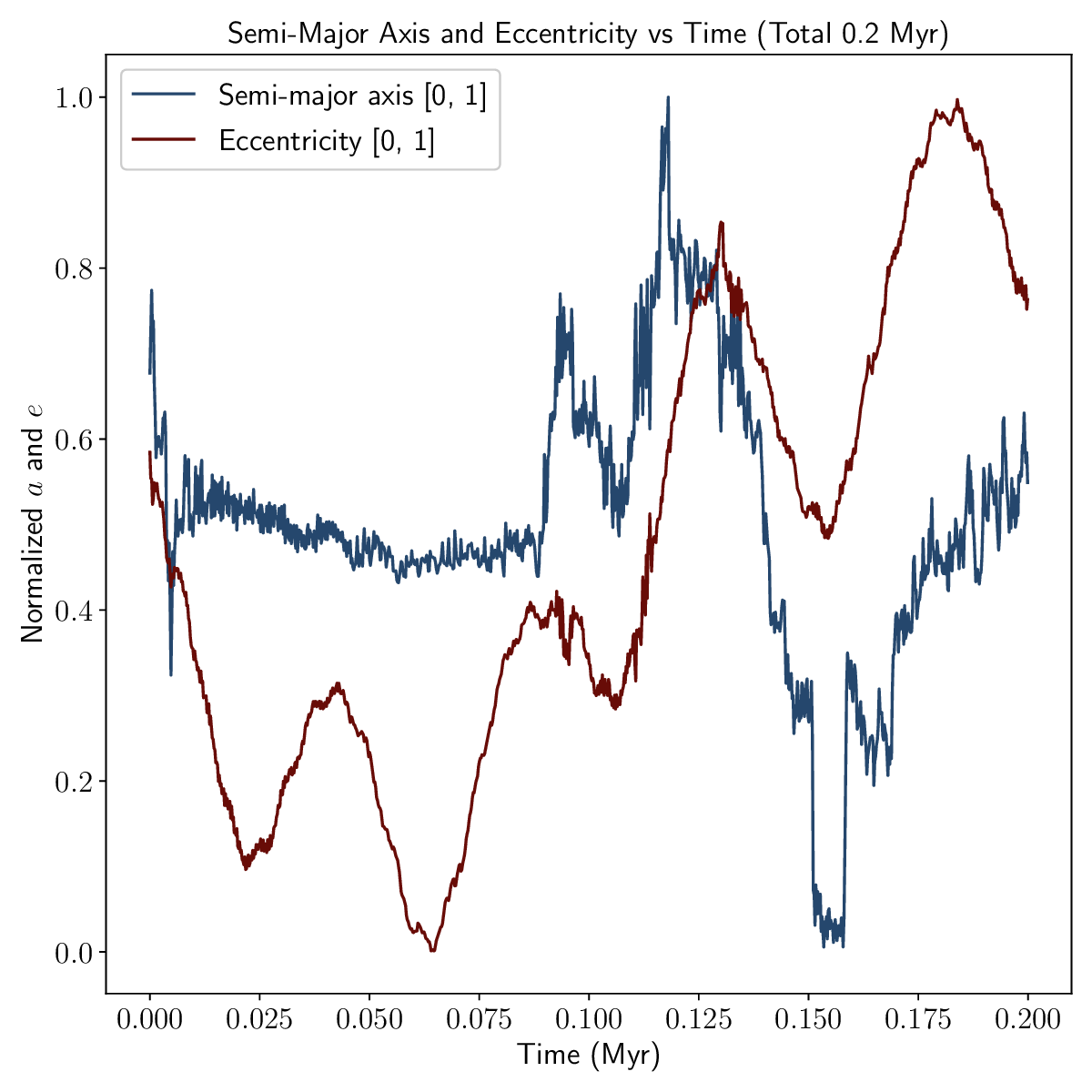}
        \vspace{0.5ex}
        (a)
    \end{minipage}
\end{minipage}\hspace{1.0cm}
\begin{minipage}{0.44\textwidth}
    \centering
    \begin{minipage}{0.9\linewidth}
        \centering
        \includegraphics[width=\linewidth]{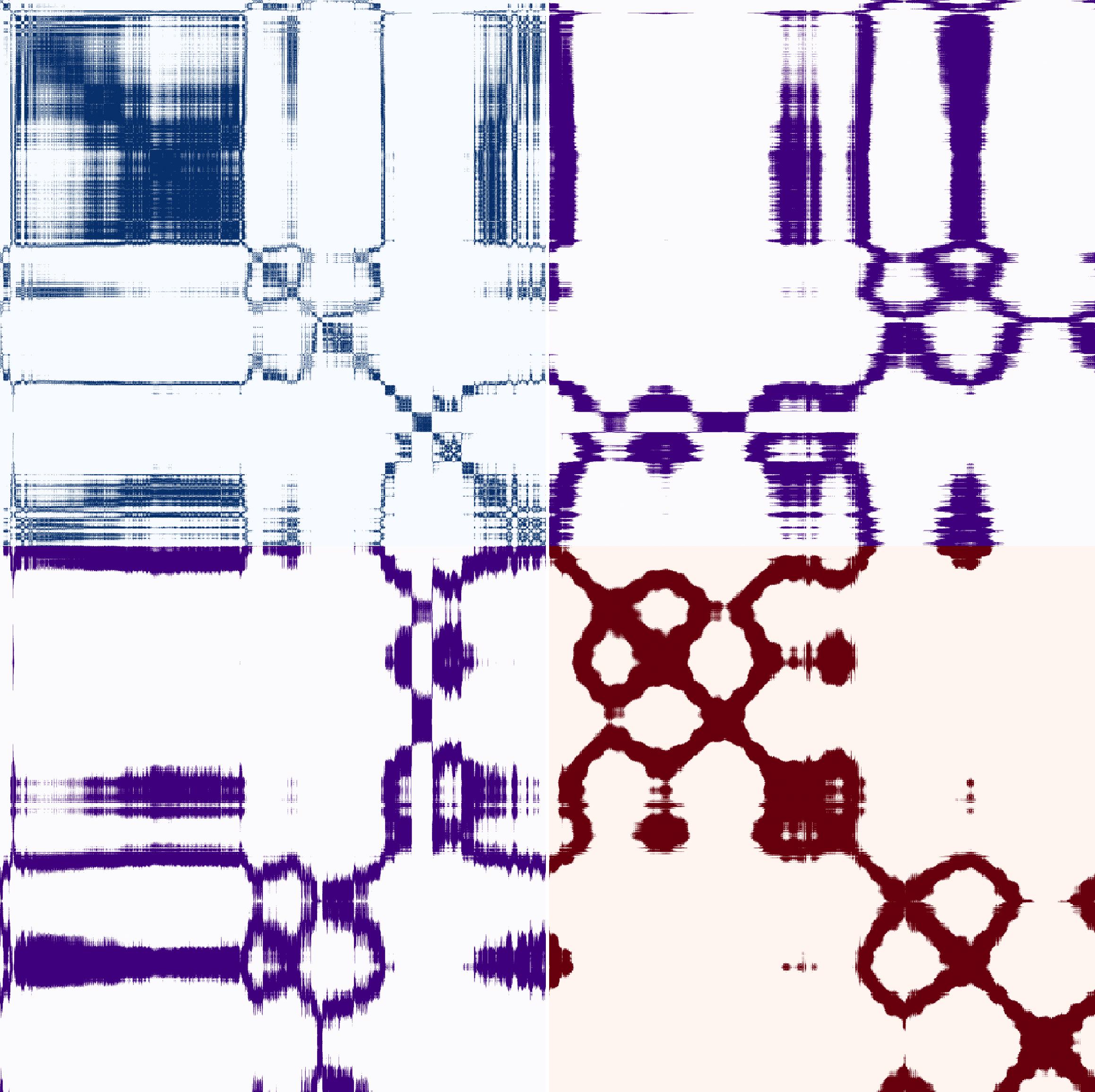}
        \vspace{0.5ex}
        (b)
    \end{minipage}
\end{minipage}

\vspace{0.2cm}

\begin{minipage}{0.44\textwidth}
    \centering
    \begin{minipage}{0.9\linewidth}
        \centering
        \includegraphics[width=\linewidth]{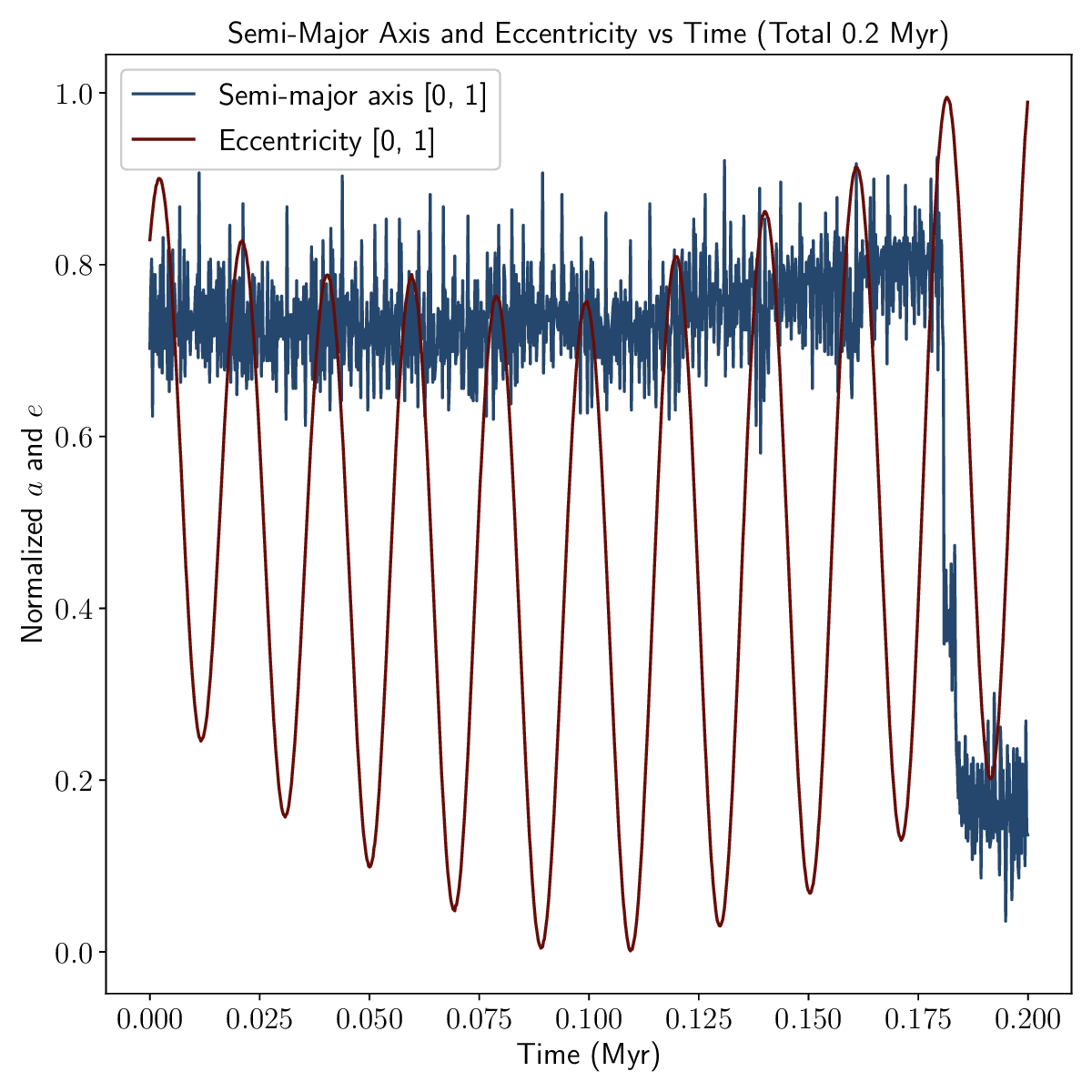}
        \vspace{0.5ex}
        (c)
    \end{minipage}
\end{minipage}\hspace{1.0cm}
\begin{minipage}{0.44\textwidth}
    \centering
    \begin{minipage}{0.9\linewidth}
        \centering
        \includegraphics[width=\linewidth]{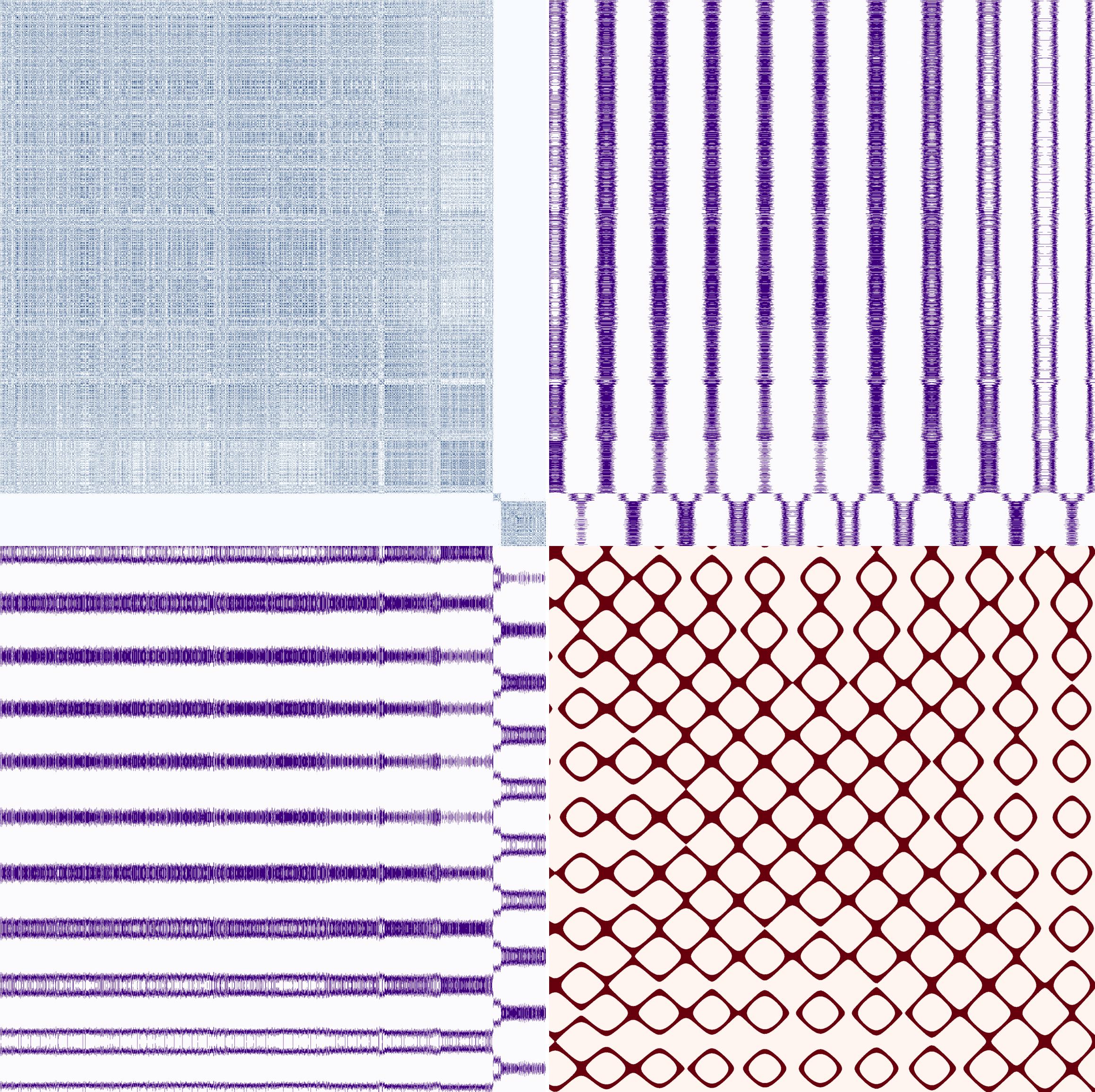}
        \vspace{0.5ex}
        (d)
    \end{minipage}
\end{minipage}

\caption{Comparative orbital evolution and recurrence structure of asteroids 2007~VP243 and 2005~WE55 over a $0.2$~Myr interval.
(a) Semi-major axis and eccentricity evolution of the dynamically unstable asteroid 2007~VP243, showing large semi-major axis excursions and irregular eccentricity growth indicative of chaotic diffusion.
(b) RP of 2007~VP243, exhibiting fragmented, non-periodic structures characteristic of chaotic, escape-prone dynamics.
(c) Semi-major axis and eccentricity evolution of the long-lived asteroid 2005~WE55, which displays regular oscillations and minimal secular drift.
(d) RP of 2005~WE55, showing ordered, grid-like patterns consistent with quasi-periodic motion.
This figure demonstrates how recurrence analysis distinguishes chaotic evolutionary pathways from stable, long-lived asteroid orbits. Here, ``normalized $a$ and $e$" denotes the semi-major axis $a$ and eccentricity $e$ time series scaled independently to the interval $[0,1]$ using min--max normalization.}
\label{fig:recurrence_plots}
\end{figure*}

RPs encode instances when a system's state revisits previous states and are constructed from a recurrence matrix \citep{thiel2004estimation, robinson2009recurrences}:
\begin{equation}
	R_{i,j} = \Theta\!\left(\epsilon - \left\| x_{i} - x_{j} \right\|\right),
\end{equation}
\noindent where $x_{i}$ and $x_{j}$ are state vectors at times $i$ and $j$, $\|\cdot\|$ is the chosen norm (here Euclidean), $\epsilon$ is a neighborhood threshold, and $\Theta$ denotes the Heaviside step function. A value $R_{i,j}=1$ indicates $\|x_i-x_j\|\le\epsilon$ (recurrence), otherwise $R_{i,j}=0$.

For this study, RPs are generated from asteroid phase-space trajectories constructed from the semi-major axis and eccentricity time series over a 0.2~Myr interval. The time series are downsampled and concatenated sequentially (semi-major axis followed by eccentricity), yielding a fixed input length of $L=4000$ samples. Recurrence matrices are computed using the \texttt{RecurrencePlot} transformer in \texttt{pyts} \citep{faouzi2020pyts}, adopting a pointwise thresholding scheme in which the recurrence threshold is set such that 20\% of pairwise distances are classified as recurrent for each time series. The resulting binary recurrence matrices are resized to $150\times150$ pixels and used as inputs to convolutional neural networks for downstream classification \citep{hatami2018classification}.

Patterns in the RPs provide diagnostics of dynamical behaviour \citep{webber2005recurrence, marwan2007recurrence}. Prominent diagonal lines are characteristic of deterministic or periodic motion (quasi-stable recurrence of orbital states); vertical or horizontal line segments are consistent with intermittent trapping (resonant behaviour); isolated points or scattered structure indicate chaotic evolution. Fractal or clustered motifs often mark dynamical transitions, such as the onset of strong perturbations or pathways toward ejection. Analysis of these features therefore aids discrimination between long-lived, diffusing, and escaping orbital trajectories.

As described in Section~\ref{sec:rp_generation}, the time series were normalized and concatenated before RP construction. This ordering produces four natural submatrices in the RPs: the upper-left and lower-right blocks correspond to self-recurrences of the semi-major axis and eccentricity, respectively, while the off-diagonal blocks represent their cross-recurrences. The time series shown in Figures~\ref{fig:recurrence_plots}a and \ref{fig:recurrence_plots}b are normalized accordingly. Colour is used in Figure~\ref{fig:recurrence_plots} solely for visualization; all CNN training and inference are performed on grayscale (binary) RPs.

Figures~\ref{fig:recurrence_plots}a and \ref{fig:recurrence_plots}b illustrate the orbital evolution and corresponding RP for asteroid 2007~VP243, which exhibits dynamical behaviour consistent with eventual ejection from the Solar System. In Figure~\ref{fig:recurrence_plots}a, the semi-major axis (blue curve) exhibits significant variability, particularly after approximately 0.1 Myr, indicating large orbital perturbations. The eccentricity (red curve) similarly shows pronounced semi-periodic oscillations and irregular growth over time. These behaviours are reflected in the RP shown in Figure~\ref{fig:recurrence_plots}b, where the upper-left submatrix (semi-major axis vs. semi-major axis) displays irregular, box-like patterns suggestive of a departure from regular orbital behaviour. The lower-right submatrix (eccentricity vs. eccentricity) also shows complex, chaotic structures consistent with the semi-periodic and unstable evolution of the eccentricity. Moreover, the cross-recurrence submatrices in purple colour (upper-right and lower-left) exhibit fragmented and dispersed patterns, highlighting the weak and erratic coupling between the semi-major axis and eccentricity. Collectively, the irregularities across all submatrices confirm the underlying dynamical instability of asteroid 2007~VP243, supporting its eventual ejection.

In contrast, Figures~\ref{fig:recurrence_plots}c and \ref{fig:recurrence_plots}d depict the orbital evolution and RP for asteroid 2005~WE55, which exhibits long-term dynamical stability and remains gravitationally bound over the integration interval. The semi-major axis evolution in Figure~\ref{fig:recurrence_plots}c remains remarkably steady for most of the integration period, with only minor fluctuations toward the end of the integration interval. The eccentricity exhibits regular, periodic oscillations throughout the entire timespan. Correspondingly, the RP in Figure~\ref{fig:recurrence_plots}d exhibits highly ordered and structured patterns. The upper-left submatrix shows densely packed horizontal and vertical lines, indicative of quasi-constant semi-major axis values interspersed with small, regular oscillations. The lower-right submatrix, representing the recurrence of eccentricity values, reveals a clear grid-like structure characteristic of stable periodic motion. Furthermore, the cross-recurrence submatrices between the semi-major axis and eccentricity display orderly, repeating motifs, signifying a consistent relationship between these orbital elements. These coherent and structured recurrence patterns affirm the dynamical stability of asteroid 2005~WE55 over long timescales.

Together, these examples demonstrate that RP morphology provides a clear diagnostic for distinguishing dynamically unstable, escape prone orbits from long-term stable asteroid trajectories.

\section{Machine Learning and Deep Learning Frameworks}
\label{sec:methods}

\begin{figure*}
    \centering
    \hspace*{-1mm} 
    \includegraphics[width=.71\textwidth]{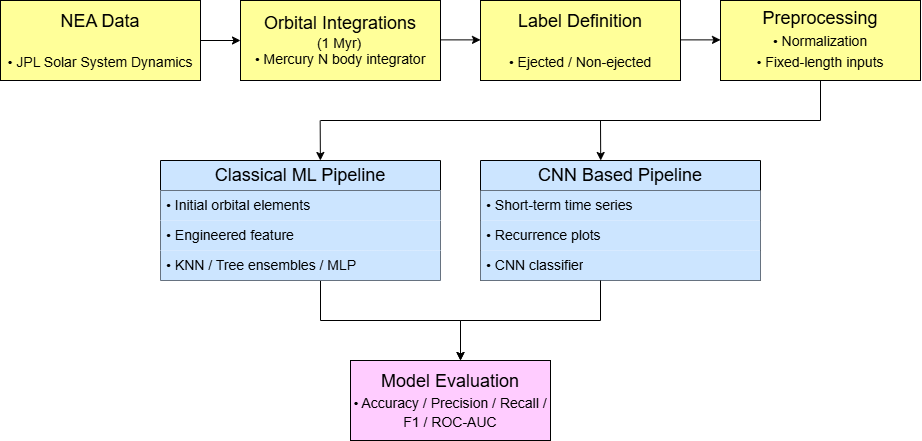}
    \caption{Traditional ML models and CNN-based pipeline.}
    \label{fig:Model Structure}
\end{figure*}

Building upon the earlier study by \cite{bora2024temporal}, which employs traditional ML models such as KNNs, Decision Trees, and ensemble-based classifiers to evaluate NEA classification and hazard assessment, this study advances the methodology by introducing a DL framework centered on CNNs. While prior efforts demonstrate the effectiveness of ML models when trained on structured orbital elements, the present study focuses on leveraging RPs derived from short-term dynamical trajectories as image-based inputs to CNNs. This approach enables the extraction of nonlinear dynamical features critical for predicting long-term asteroid ejection. To provide a comprehensive evaluation, the previously employed ML models are retained as baselines, and the overall modeling framework encompassing both traditional ML and the proposed CNN-based pipeline is depicted in Figure~\ref{fig:Model Structure}.

A comparative analysis between these ML models and the proposed DL models is performed to evaluate improvements in predictive performance.

\subsection{Machine Learning Models}

The ML models selected for this study are chosen based on their proven effectiveness in structured data classification tasks \citep{wang2020systematic}. They are recognized for their capacity to manage complex decision boundaries and to generalize effectively across diverse datasets.

(i) KNN: A non-parametric algorithm that assigns class labels by evaluating the closeness of a specified number of neighboring points in the feature space, with its performance largely influenced by the choice of neighbors and distance metric.

(ii) Decision Tree: A model that partitions the dataset through recursive splits on feature thresholds, creating a structured set of decision rules. Its interpretability makes it valuable for identifying and understanding classification patterns.

(iii) RF: An ensemble learning technique that builds multiple decision trees and combines their predictions to improve robustness and minimize overfitting.

(iv) ExtraTree: Similar to RF but introduces greater randomness in feature selection and threshold determination, enhancing generalization while maintaining computational efficiency.

(v) MLP: A neural network with multiple hidden layers that models complex, non-linear patterns in data, where the choice of activation functions and weight optimization significantly influences performance.

(vi) AdaBoost: A boosting method that iteratively trains weak classifiers, increasing the weights of misclassified instances to enhance overall predictive accuracy.

(vii) GB: A boosting algorithm that sequentially builds decision trees, optimizing for the reduction of residual errors at each step.

These models represent a diverse set of classification techniques, spanning instance-based learning (KNN) \citep{aha1991instance}, decision tree ensembles (RF, ExtraTree, AdaBoost, GB) \citep{dietterich2000ensemble}, and neural networks (MLP). The next section outlines the dataset preparation, feature selection, and training methodology used to evaluate their performance. All these ML models are implemented using the scikit-learn library in Python \citep{pedregosa2011scikit}.

\subsection{Implementation and Training of Machine Learning Models}

The dataset used for ML classification was derived from the numerically integrated NEA population described in Section~\ref{sec:data_preprocessing}. Following the augmentation and preprocessing procedures outlined in Section~\ref{sec:augmentation}, a balanced dataset of 15,000 objects per class was constructed for ML training. Classification is based on orbital parameters extracted at the initial epoch, including semi-major axis, eccentricity, inclination, argument of perihelion, longitude of the ascending node, mean anomaly, perihelion distance, and NEA dynamical class (Atens, Apollos, Amors, or Atiras).

An 80--20 train--test split with equal class representation is adopted to ensure unbiased learning. Hyperparameter optimization is performed exclusively on the training data, while model performance is evaluated on the held-out test set using accuracy, precision, recall, F1 score, and confusion matrices. These ML models provide a robust baseline for comparison with the CNN-based approaches introduced in subsequent subsections.

\subsection{Convolutional Neural Network}
\label{sec:cnn}

RPs constructed from short-term orbital time series are classified using a CNN. The CNN operates on grayscale RP images resized to $150 \times 150$ pixels and is employed as a pattern-recognition tool to extract spatial signatures associated with stable and unstable dynamical evolution. The CNN captures spatial patterns in RPs, such as filamentary structures and clustered regions that are indicative of distinct dynamical regimes.

The CNN is implemented using \texttt{Keras} with a \texttt{TensorFlow} backend \citep{gulli2017deep,abadi2016tensorflow}, following standard practice in image-based time-series classification \citep{lecun1998convolutional,krizhevsky2012imagenet,hatami2018classification}. Architectural choices and training parameters were selected through systematic hyperparameter tuning (Subsection~\ref{sec:hyperparam}).

\subsection{CNN Model Architecture and Training}
\label{sec:cnn_arch}

\begin{figure}
	\centering
	\includegraphics[width=0.60\textwidth]{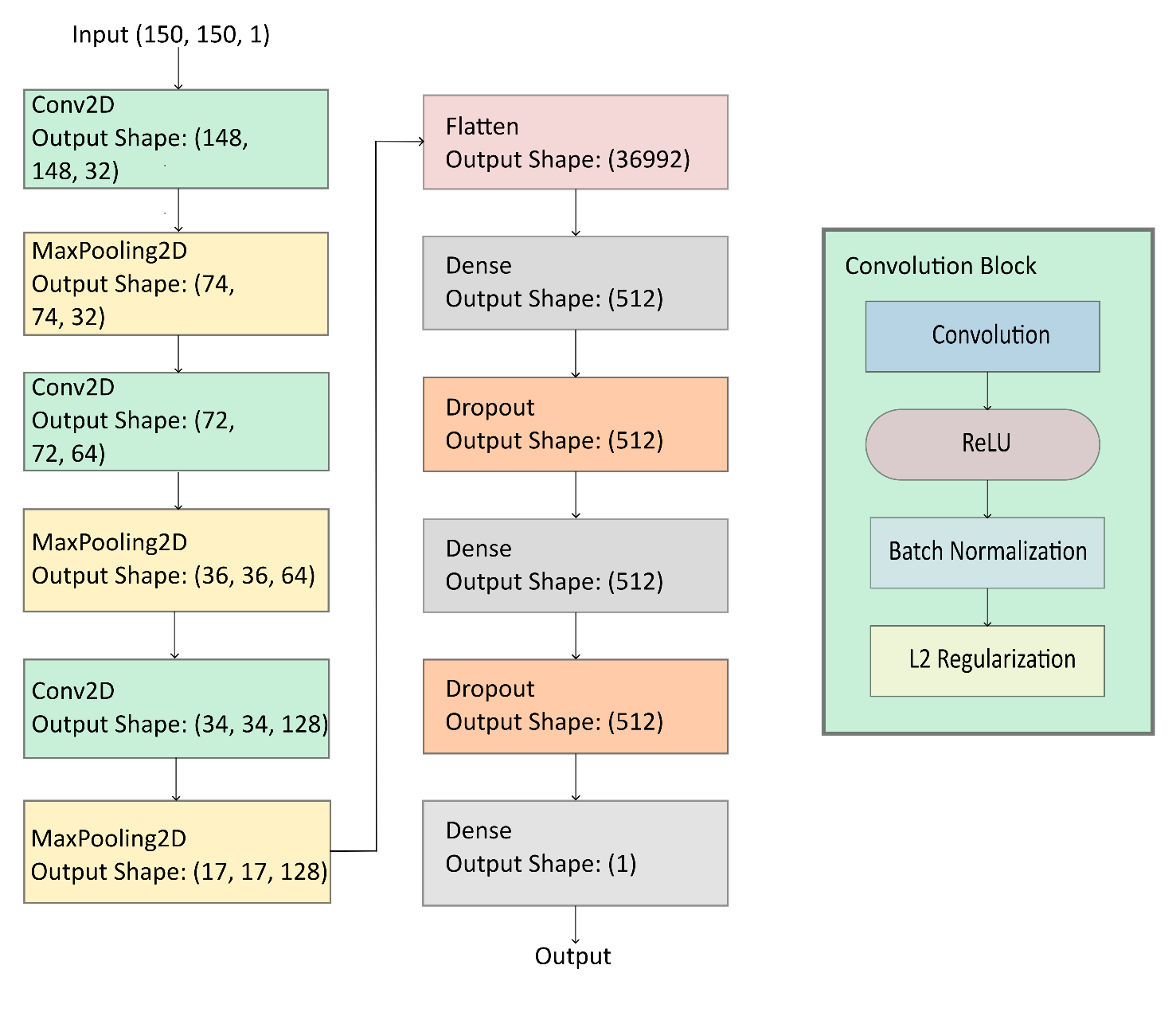}
	\caption{CNN architecture used for classification of RPs.}
	\label{fig:CNN_Model_Architecture}
\end{figure}

The CNN architecture, illustrated in Figure~\ref{fig:CNN_Model_Architecture}, consists of three convolutional layers with 32, 64, and 128 filters, respectively, followed by max-pooling layers, and two fully connected layers with 512 neurons each. The convolutional kernel size is treated as a hyperparameter and optimised during training. A single sigmoid-activated neuron forms the output layer for binary classification. ReLU activations are used throughout. L2 regularization is applied to both convolutional and dense layers, and dropout is applied after the dense layers to mitigate overfitting. The specific regularization strengths and dropout rates are selected through hyperparameter tuning, as detailed in Subsection~\ref{sec:hyperparam} and Table~\ref{tab:hyperparameter_search}.

In the context of neural network training, the term \emph{epoch} refers to one complete pass through the training dataset. This usage is distinct from the \emph{astronomical epoch}, which denotes the reference time at which orbital elements are specified.

A three-layer convolutional structure was adopted after empirical testing of deeper architectures. Preliminary experiments with four- and five-layer CNN architectures did not improve classification accuracy and instead resulted in increased overfitting during training. In contrast, the three-layer configuration consistently achieved stable convergence and robust validation performance across a wide range of hyperparameter settings. The chosen architecture therefore provides sufficient representational capacity for recurrence plot images while avoiding unnecessary model complexity.

Model training employs the Adam optimiser \citep{kingma2014adam} with binary cross-entropy loss. The dataset is divided into training, validation, and test sets in an 80:10:10 ratio, comprising 29,233 training images, 3,502 validation images, and 3,502 test images. This explicit validation split is required for CNN training to enable early stopping and reliable hyperparameter selection.

To ensure stable convergence and avoid overfitting, EarlyStopping, ReduceLROnPlateau, and ModelCheckpoint callbacks were used during training \citep{prechelt2002early,smith2017cyclical}. The final model is evaluated exclusively on the held-out test set, and its performance metrics are presented in Section~\ref{sec:classification_results}.

\subsection{Hyperparameter Tuning}
\label{sec:hyperparam}

Hyperparameters for both the classical ML models and the CNN were systematically optimized to ensure robust generalization and to avoid overfitting \citep{hastie2009elements}. For the classical ML models, hyperparameter tuning was performed using grid search combined with 5-fold cross-validation on the training set. The optimal configuration for each model was selected based on the mean cross-validation accuracy. The principal hyperparameters explored for each ML model are summarized in Table~\ref{tab:ml_hyperparams}.

\begin{table}
	\centering
	\begin{tabular}{|l|l|}
		\hline
		\textbf{Model} & \textbf{Key hyperparameters tuned} \\
		\hline
		KNN & \texttt{n\_neighbors}, \texttt{weights} \\
		Decision Tree & \texttt{max\_depth}, \texttt{min\_samples\_split} \\
		RF/ExtraTree & \texttt{n\_estimators}, \texttt{max\_features} \\
		AdaBoost/GB & \texttt{n\_estimators}, \texttt{learning\_rate} \\
		MLP & \texttt{hidden\_layer\_sizes}, \texttt{alpha} (L2) \\
		\hline
	\end{tabular}
	\caption{Representative hyperparameters tuned for the classical ML models.}
	\label{tab:ml_hyperparams}
\end{table}

\begin{table}
	\centering
	\begin{tabular}{|c|c|}
		\hline
		\textbf{Hyperparameter} & \textbf{Values} \\
		\hline
		Learning Rates & [0.0001, \textbf{0.001}, 0.01, 0.1] \\
		\hline
		Batch Sizes & [32, 64, \textbf{128}, 256] \\
		\hline
		Kernel Sizes & [\textbf{(3, 3)}, (5, 5)] \\
		\hline
		Dropout Rates & [0.3, 0.4, \textbf{0.5}, 0.6, 0.7] \\
		\hline
		L2 Regularization (Conv layer) & [0.0009, 0.0095, \textbf{0.001}, 0.0015, 0.002, 0.0025, 0.003] \\
		\hline
		L2 Regularization (Dense layer) & [0.0009, 0.0095, 0.001, 0.0015, \textbf{0.002}, 0.0025, 0.003] \\
		\hline
		Number of Epochs & [15, \textbf{20}, 25, 30] \\
		\hline
	\end{tabular}
	\caption{Hyperparameter ranges explored during CNN training. Values highlighted in bold correspond to the configuration yielding the best validation performance.}
	\label{tab:hyperparameter_search}
\end{table}

For the CNN, hyperparameters governing optimization, regularization, and architecture were explored through a grid-based search over predefined ranges, summarized in Table~\ref{tab:hyperparameter_search}. Model selection was performed using the validation set, where the optimal hyperparameter configuration was chosen based on the highest validation accuracy across trials, while within each training run the best model weights were selected according to the minimum validation loss. Early stopping and learning-rate scheduling were employed to stabilize convergence and prevent overfitting.

\subsection{Model Evaluation Metrics}

Model performance is assessed using standard classification metrics, including accuracy, precision, recall, F1 score, and confusion matrices. To provide threshold-independent evaluation, receiver operating characteristic (ROC) curves \citep{hanley1982meaning} and precision–recall (PR) curves \citep{davis2006relationship} are also computed, together with their corresponding areas under the curve (AUC).

ROC curves quantify overall discriminative capability, while PR curves are particularly informative for evaluating the balance between false positives and false negatives. Together, these complementary metrics provide a robust assessment of classification performance across both ML and DL models.

\subsection{Computational Environment and Reproducibility}

All numerical integrations and machine-learning experiments were conducted using Python~3.8. Classical data processing and analysis employed NumPy~1.21.1, pandas~1.4.3, and matplotlib~3.5.0. Deep-learning models were implemented using Keras~2.10.0 with a TensorFlow~2.10.1 backend, compiled with CUDA support. Model training was performed on the Bhaskara high-performance computing system [NVIDIA DGX-1 equipped with $8\times$ Tesla V100 GPUs] at IIT (ISM) Dhanbad, using NVIDIA driver version~525.60.13 and CUDA~12.0. A fixed random seed of 42 was applied consistently across the Python training pipeline (Python \texttt{random}, NumPy, and TensorFlow RNGs) using \texttt{random.seed(42)}, \texttt{numpy.random.seed(42)}, and \texttt{tf.random.set\_seed(42)}. While this controls the main stochastic elements of training (e.g., weight initialization and data shuffling), exact bitwise reproducibility cannot be guaranteed due to minor non-deterministic GPU operations; however, repeated runs produce broadly consistent results.

\section{Classification Results}
\label{sec:classification_results}

To classify long-term 1-Myr ejection outcomes of NEAs, we evaluated the classification performance of traditional ML models and CNNs across distinct feature sets. Specifically, we compare models trained on initial orbital parameters to those trained on trajectory-derived features, assessing their respective classification performance and generalization for finite-time, ensemble-level outcome labels.

\subsection{Classification with Initial Orbital Parameters}

\begin{figure*}
	\centering
	\hspace*{-1mm} 
	\includegraphics[width=0.64\textwidth]{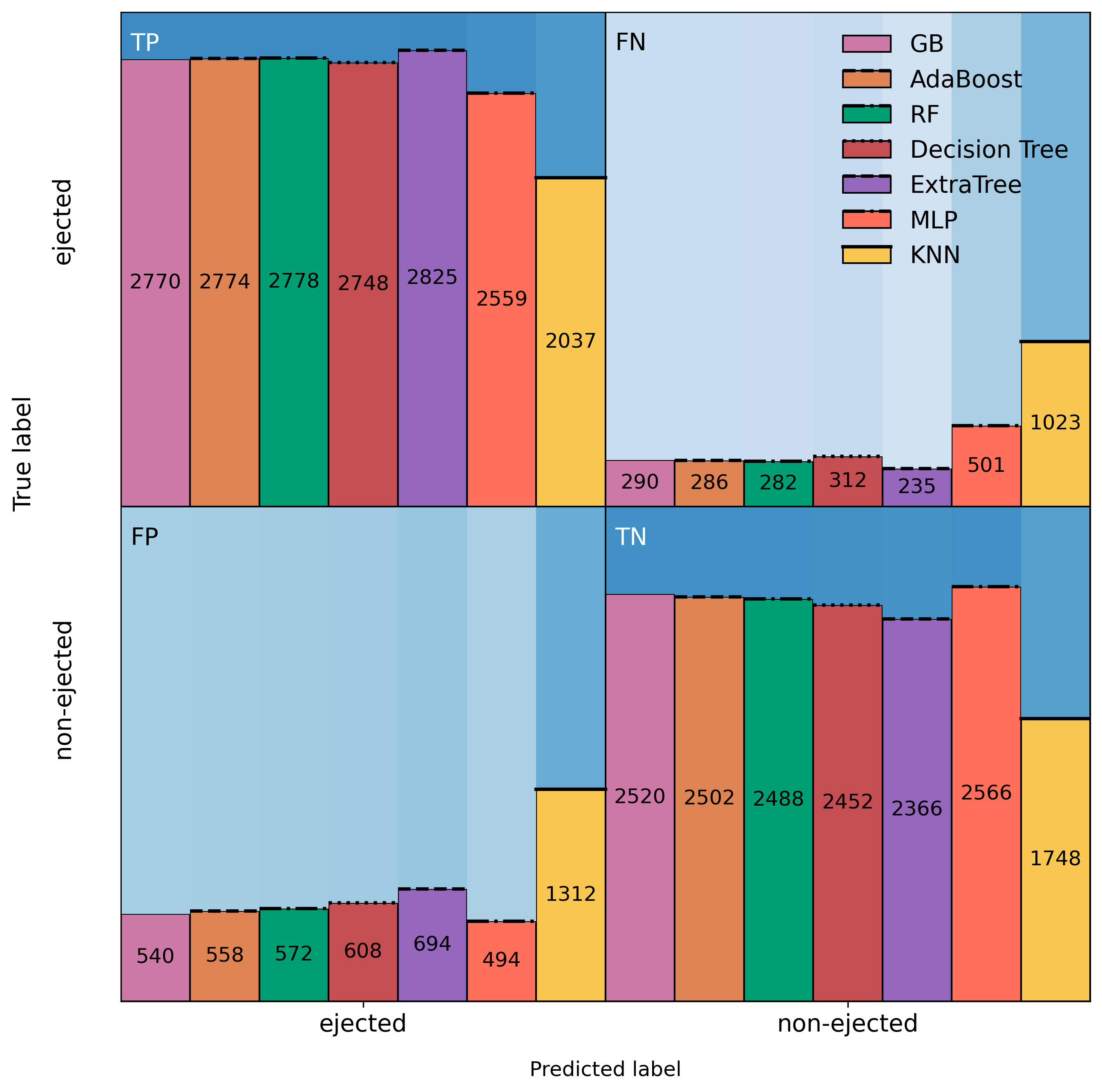}
	\caption{Confusion matrices for seven classification models shown in a unified 2×2 layout separating true positives (TP), false negatives (FN), false positives (FP), and true negatives (TN). Coloured bars denote individual models, ordered by decreasing overall performance from left to right, with bar heights normalized by the total number of objects in the corresponding true class. Numerical annotations indicate absolute counts in each cell.}
	\label{fig:confusion_matrix_global}
\end{figure*}

\begin{table}
\centering
\begin{tabular}{|l|c|c|c|c|}
\hline
\textbf{Model} & \textbf{Accuracy} & \textbf{Recall} & \textbf{Precision} & \textbf{F1 Score} \\
\hline
KNN & 0.6185 & 0.6657 & 0.6082 & 0.6357 \\
Decision Tree & 0.8497 & 0.8980 & 0.8188 & 0.8566 \\
\textbf{RF} & 0.8605 & 0.9078 & 0.8293 & 0.8668 \\
ExtraTree & 0.8482 & 0.9232 & 0.8028 & 0.8588 \\
MLP & 0.8374 & 0.8363 & 0.8382 & 0.8372 \\
\textbf{AdaBoost} & 0.8621 & 0.9065 & 0.8325 & 0.8680 \\
\textbf{GB} & 0.8644 & 0.9052 & 0.8369 & 0.8697 \\
\hline
\end{tabular}
\caption{Evaluation metrics for traditional ML models.}
\label{tab:ML_Models_Performance}
\end{table}

This subsection evaluates the performance of seven ML models trained solely on the initial orbital parameters of NEAs. The evaluation relied on four key metrics: accuracy, recall, precision, and F1 score, as summarized in Table \ref{tab:ML_Models_Performance}. Restricting the input to the initial ephemerides provides insight into the predictive capabilities of ML models operating with minimal feature information, a strategy also adopted in earlier asteroid-classification studies such as \cite{mako2005classification, smirnov2017identification,carruba2020machine}.

The models tested included both tree-based and non-tree-based algorithms. Among them, tree-based models, Decision Tree, RF, ExtraTree, AdaBoost, and GB, consistently outperformed non-tree-based models such as KNN and MLP, in agreement with the findings of \cite{bahel2021supervised, bora2024temporal}. This result highlights the strength of tree-based approaches in handling non-linear relationships and providing robust predictions based solely on orbital elements.

As shown in Table \ref{tab:ML_Models_Performance}, the best-performing models marked in bold are RF, AdaBoost, and GB, each achieving an accuracy of 86\%, highlighting their strong classification performance. These models achieved recall values between 0.90 and 0.91, indicating strong ability to correctly identify ejected asteroids. Precision scores for these models were also quite strong (0.82–0.84), reflecting their ability to minimize false positives. The F1 scores for these models, which are approximately 0.87, further corroborate their balanced performance, with precision and recall achieving a desirable trade-off. The evaluation values are similar to the scores of \cite{vignesh2024asteroid, sajid2024machine}.

The Decision Tree model, although slightly trailing the ensemble methods, demonstrated strong performance with an accuracy of 85\%, a recall of 0.90, and an F1 score of 0.86. Its simplicity and interpretability make it particularly valuable, especially in scenarios where model transparency is crucial \citep{izza2020explaining}. The ExtraTree model yielded similar results, further emphasizing the advantage of tree-based methodologies.

Conversely, the KNN model exhibited the lowest performance, with an accuracy of 62\%. Its relatively poor recall (0.67) and precision (0.61) suggest sensitivity to the curse of dimensionality, where simple orbital parameters do not form clearly defined clusters necessary for KNN to be effective. Consistent with our findings, their analysis likewise reported KNN as the weakest-performing model \citep{chandra2025enhanced}.

The MLP model achieved an accuracy of 84\% and an F1 score of 0.84, demonstrating competitive results relative to the tree-based models. However, despite its increased representational capacity, the MLP did not outperform the simpler ensemble-based approaches, indicating that for datasets restricted to ephemerides, tree ensembles remain the more effective choice. Notably, \citet{almousa2025asteroid} reported exceptionally strong results using MLPs for asteroid-group classification; however, their task is considerably more amenable to such models, as the underlying decision boundaries can, in some instances, be reduced to a near-linear separation problem unlike the substantially more complex dynamical classification problem addressed in our study.

Figure~\ref{fig:confusion_matrix_global} presents a compact, unified \(2\times2\) representation of the seven confusion matrices. The four quadrants correspond to: top-left = true positives (TP), top-right = false negatives (FN), bottom-left = false positives (FP) and bottom-right = true negatives (TN), with the horizontal axis indicating the predicted label (ejected / non-ejected) and the vertical axis the true label. Each model is shown as a coloured vertical bar inside every cell, and models are ordered left--to--right by decreasing overall performance. Bar heights indicate normalized per-class rates for the corresponding model, and numerical annotations provide the absolute counts. Behind each bar the blue-toned cell background is shaded from light to dark according to the absolute count in that cell (darker shading denotes larger sample counts, lighter shading denotes fewer), so that both relative per-class rates and the underlying sample sizes are immediately visible. The figure thus makes method-to-method contrasts explicit: the top ensembles (GB, AdaBoost, RF) display large TP and TN rates together with dark backgrounds in those cells, whereas KNN shows a reduced TP rate with comparatively darker FN and FP backgrounds (e.g.\ TP=2037, FN=1023, FP=1312), indicating its less balanced behaviour. Decision Tree, ExtraTree and MLP occupy intermediate profiles between these extremes. These visual patterns corroborate the tabulated metrics in Table~\ref{tab:ML_Models_Performance} and succinctly illustrate the superior class-wise balance achieved by the ensemble classifiers.

While our best-performing models achieved F1 scores around 0.87, slightly lower than those reported in \cite{smirnov2024comparative} for resonance classification, this comparison must be contextualized. Smirnov’s study identified mean-motion resonances within a relatively short integration period of $10^{5}$ years (0.1 Myr), whereas our task involves classifying asteroid ejection outcomes over a significantly longer duration of 1 Myr. This longer timescale introduces greater dynamical complexity, driven by cumulative gravitational perturbations and chaotic orbital evolution, making the classification task inherently more difficult. Additionally, resonance identification benefits from distinct dynamical signatures, while ejection outcomes are more diffuse and less separable. Therefore, the fact that our models still achieved balanced and high F1 scores based solely on ephemeris-derived features underscores the robustness and relevance of our ML approach for long-term dynamical predictions in asteroid studies.

To gain insight into the dynamical factors governing asteroid classification, we examined the feature importance derived from the tree-based models. Among the three best-performing classifiers, we focus on the GB model as a representative case, without loss of generality, since the inferred feature rankings are consistent across the ensemble methods. The semi-major axis emerged as the most influential parameter, contributing approximately 72\% to the overall model decision-making process. It was followed by eccentricity (13\%) and inclination (5\%), indicating that orbital size and shape are primary determinants in predicting asteroid ejection outcomes. Other features, such as perihelion distance (3\%), argument of perihelion (3\%), mean anomaly (2\%), asteroid type (1\%), and longitude of the ascending node (1\%), played comparatively smaller roles. This ordering of feature importance highlights that the initial orbital geometry is particularly critical for classification, with semi-major axis and eccentricity alone accounting for approximately 85\% of the discriminative power. This strong dependence on the semi-major axis and eccentricity is consistent with the findings of \citep{mako2005classification, smirnov2024comparative}, who similarly identified these orbital elements as the most influential factors in asteroid classification tasks.

These findings are consistent with physical expectations, as the semi-major axis and eccentricity directly influence an asteroid's interaction with planetary perturbations and resonances, which are key mechanisms driving orbital evolution and potential ejection from the Solar System \citep{morbidelli2002modern}. The relatively lower importance of angular parameters suggests that, at the initial stage, the spatial orientation of the orbit exerts less impact on long-term dynamical outcomes compared to the orbit's size and shape.

In summary, this analysis demonstrates that robust classification of NEA ejection outcomes can be achieved using tree-based ML models with only the initial orbital parameters. The strong performance of these models, despite minimal input data, suggests promising avenues for future research in asteroid dynamics and classification methodologies. Similar effectiveness of tree-based classifiers in asteroid hazard prediction was also reported by \cite{malakouti2023machine}.

\subsection{Classification with Trajectory-Based Features}

\begin{figure*}
\centering
\hspace*{-3mm}

\begin{minipage}{0.48\textwidth}
    \centering
    \includegraphics[width=0.8\linewidth]{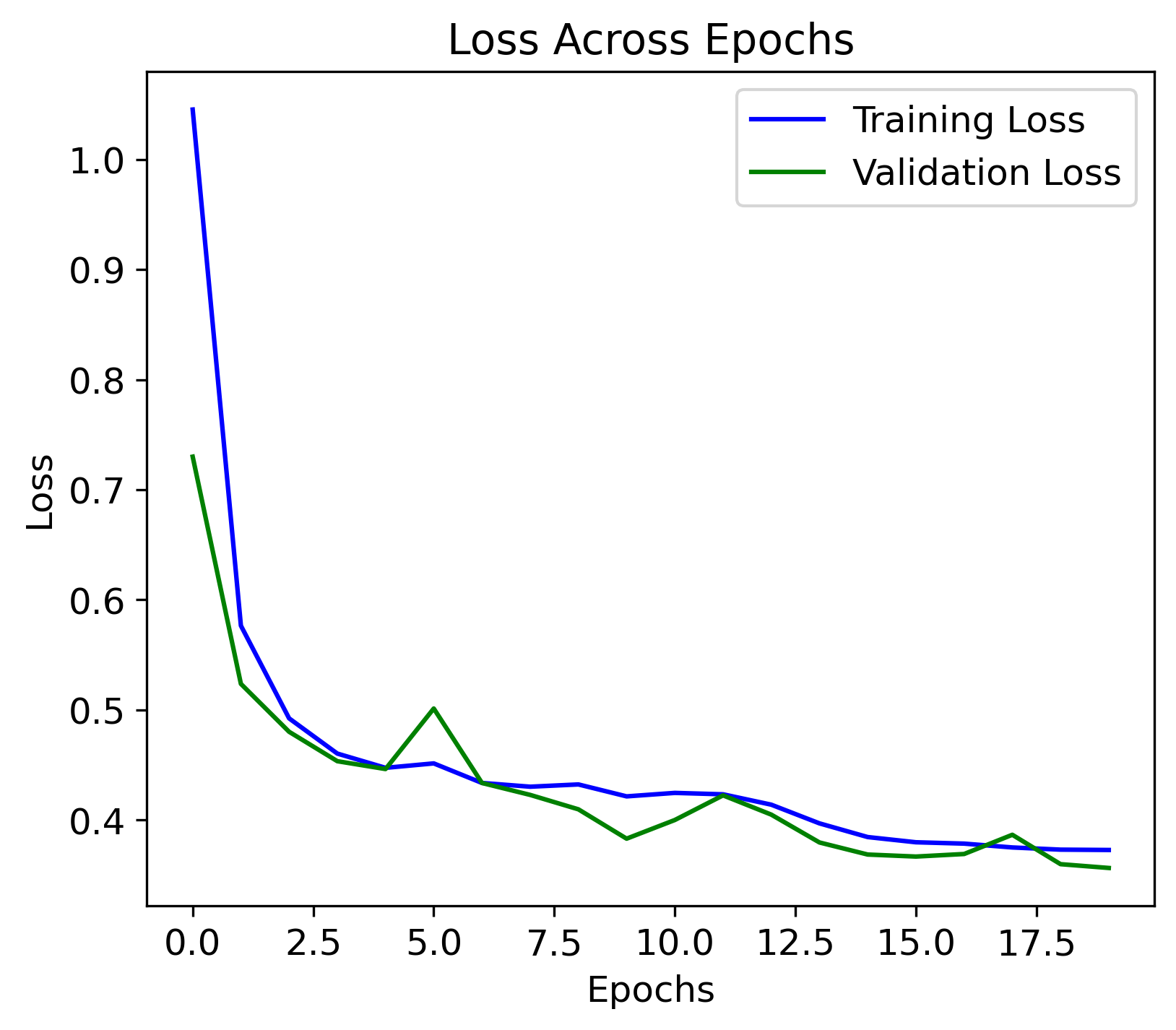}

    \vspace{0.5ex}
    (a)
\end{minipage}\hspace{0.5cm}
\begin{minipage}{0.48\textwidth}
    \centering
    \includegraphics[width=0.8\linewidth]{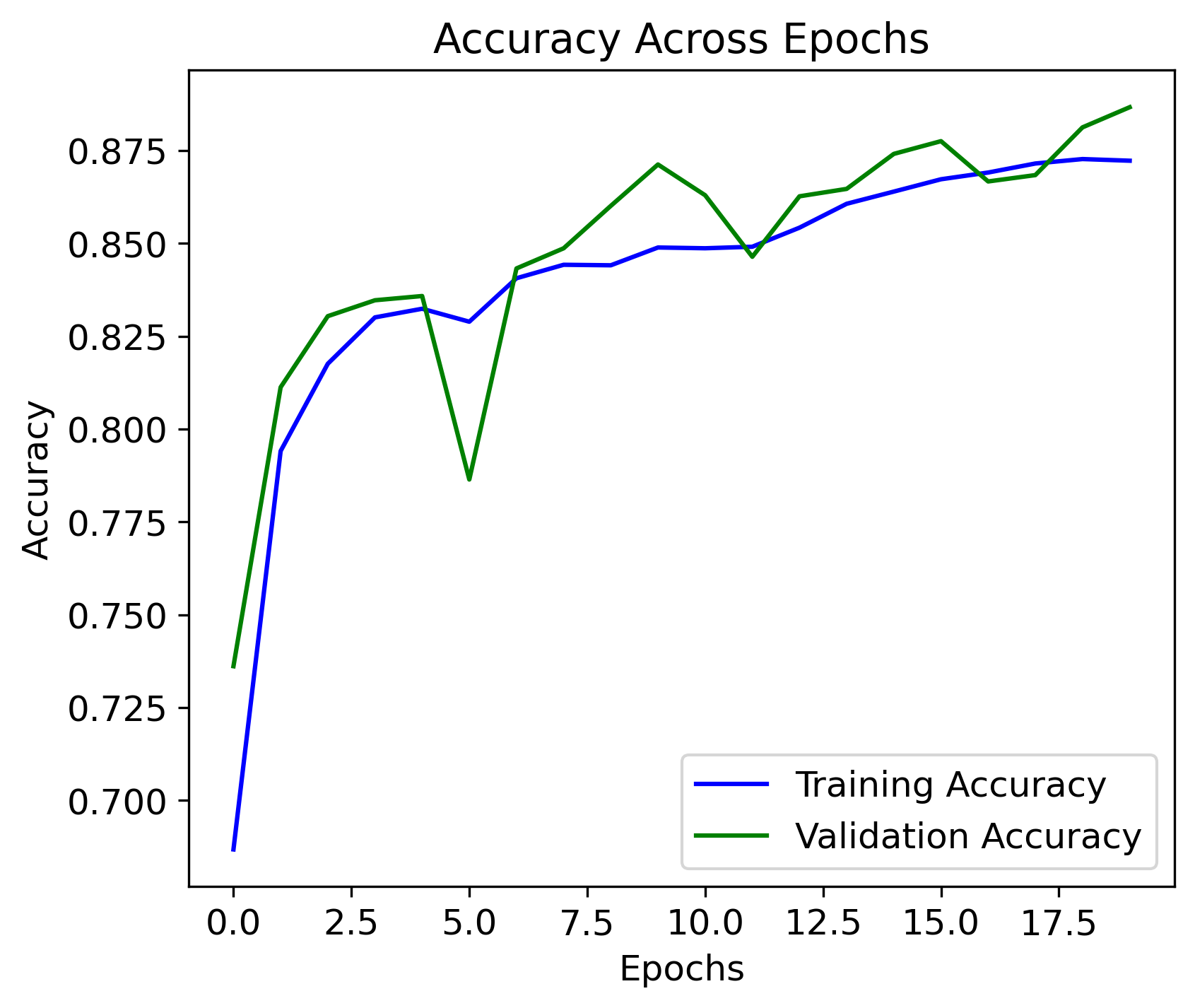}

    \vspace{0.5ex}
    (b)
\end{minipage}

\vspace{0.5cm}

\hspace*{-3mm}

\begin{minipage}{0.48\textwidth}
    \centering
    \includegraphics[width=0.8\linewidth]{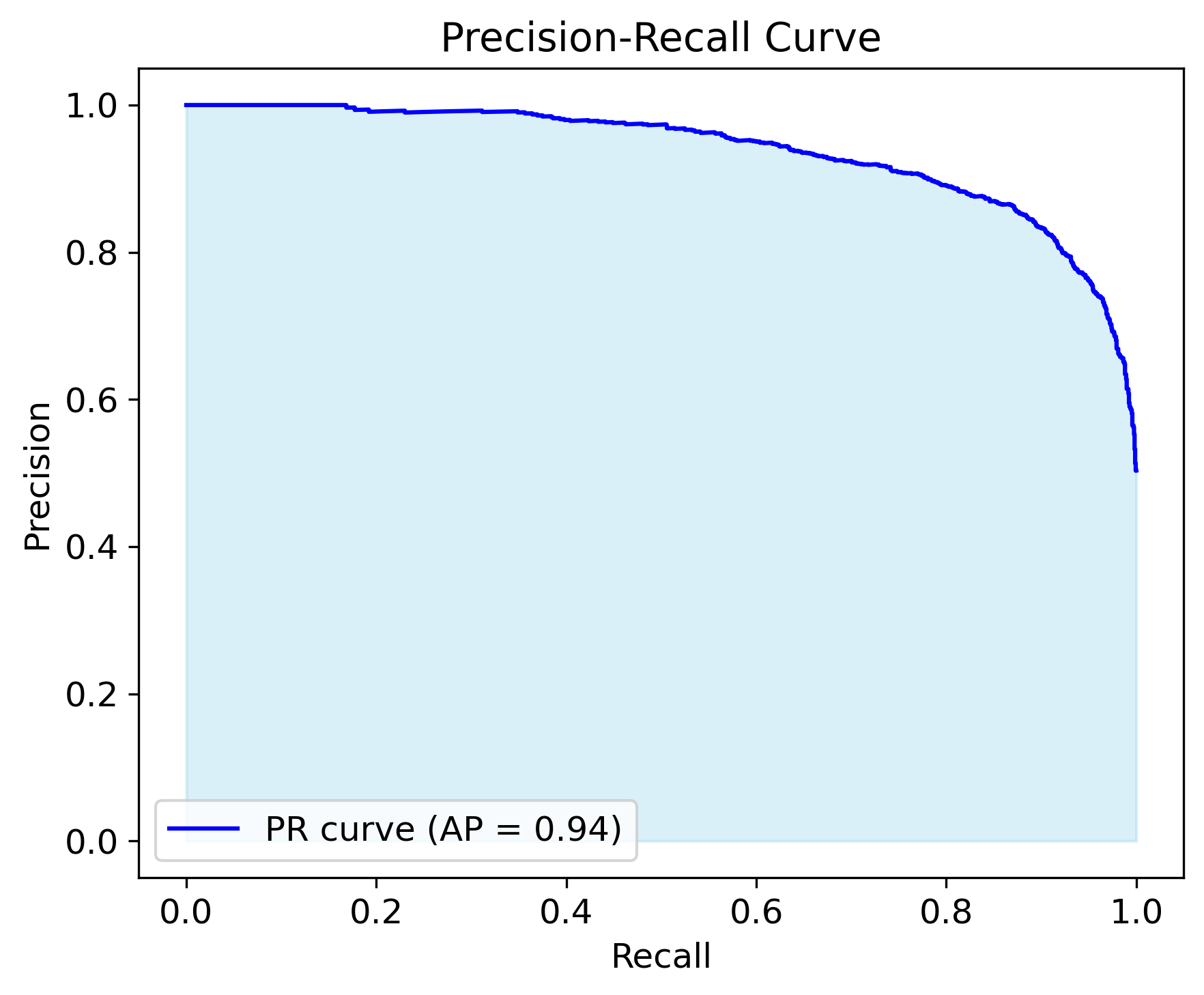}

    \vspace{0.5ex}
    (c)
\end{minipage}\hspace{0.5cm}
\begin{minipage}{0.48\textwidth}
    \centering
    \includegraphics[width=0.8\linewidth]{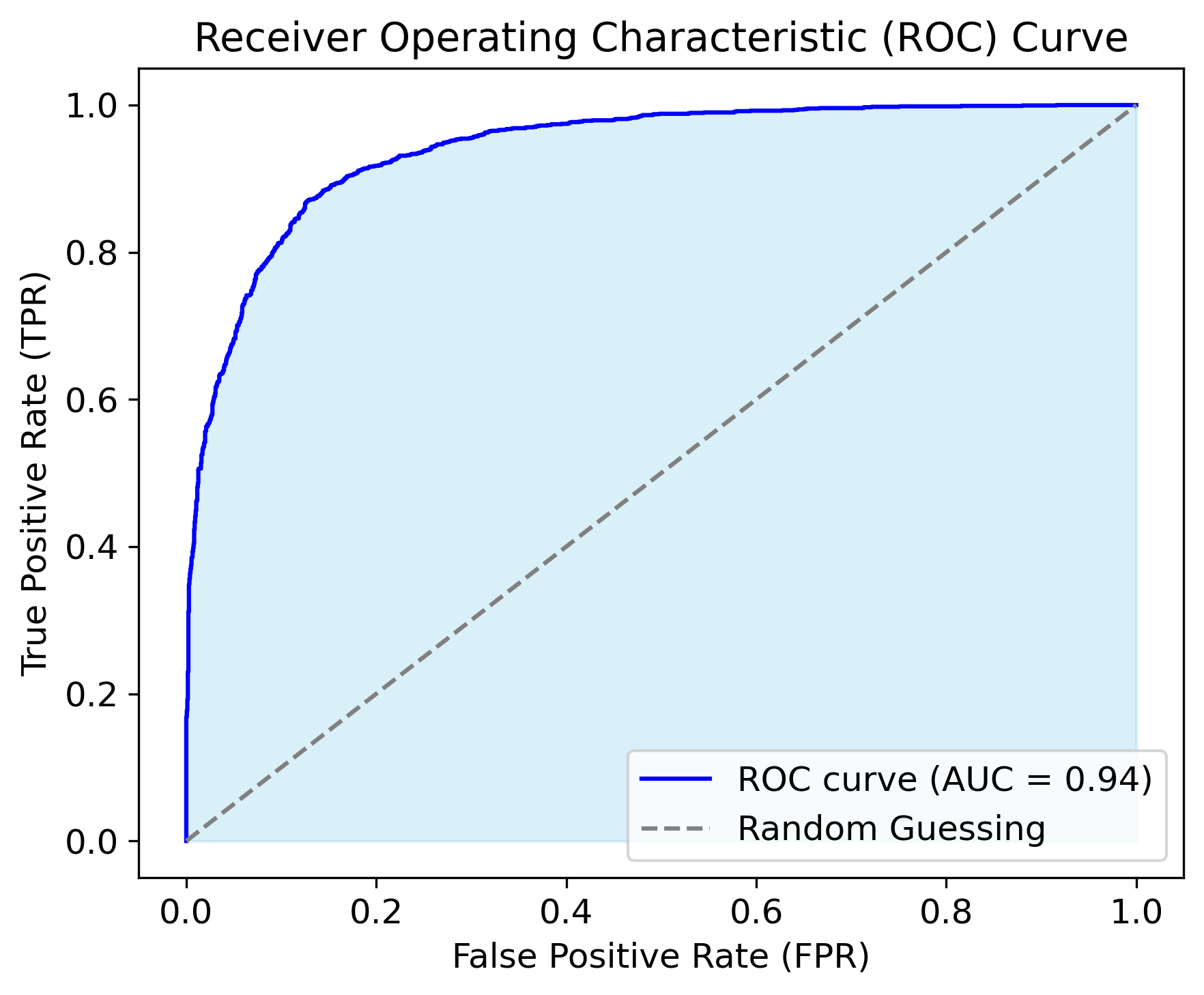}

    \vspace{0.5ex}
    (d)
\end{minipage}

\caption{Model performance summary.
(a) Training and validation loss as a function of epoch, illustrating rapid initial convergence followed by gradual stabilization.
(b) Training and validation accuracy across epochs, showing consistent improvement and convergence between the two sets.
(c) PR curve on the test set, characterizing the trade-off between precision and recall; the shaded region represents the area under the PR curve, providing a summary measure of classification performance.
(d) ROC curve on the test set, showing the true positive rate as a function of the false positive rate; the shaded region denotes the area under the ROC curve (AUC), which quantifies the overall discriminative capability of the model, while the diagonal line corresponds to random-guess performance.}
\label{fig:performance}
\end{figure*}

\begin{figure}
    \centering
    \includegraphics[width=0.4\textwidth]{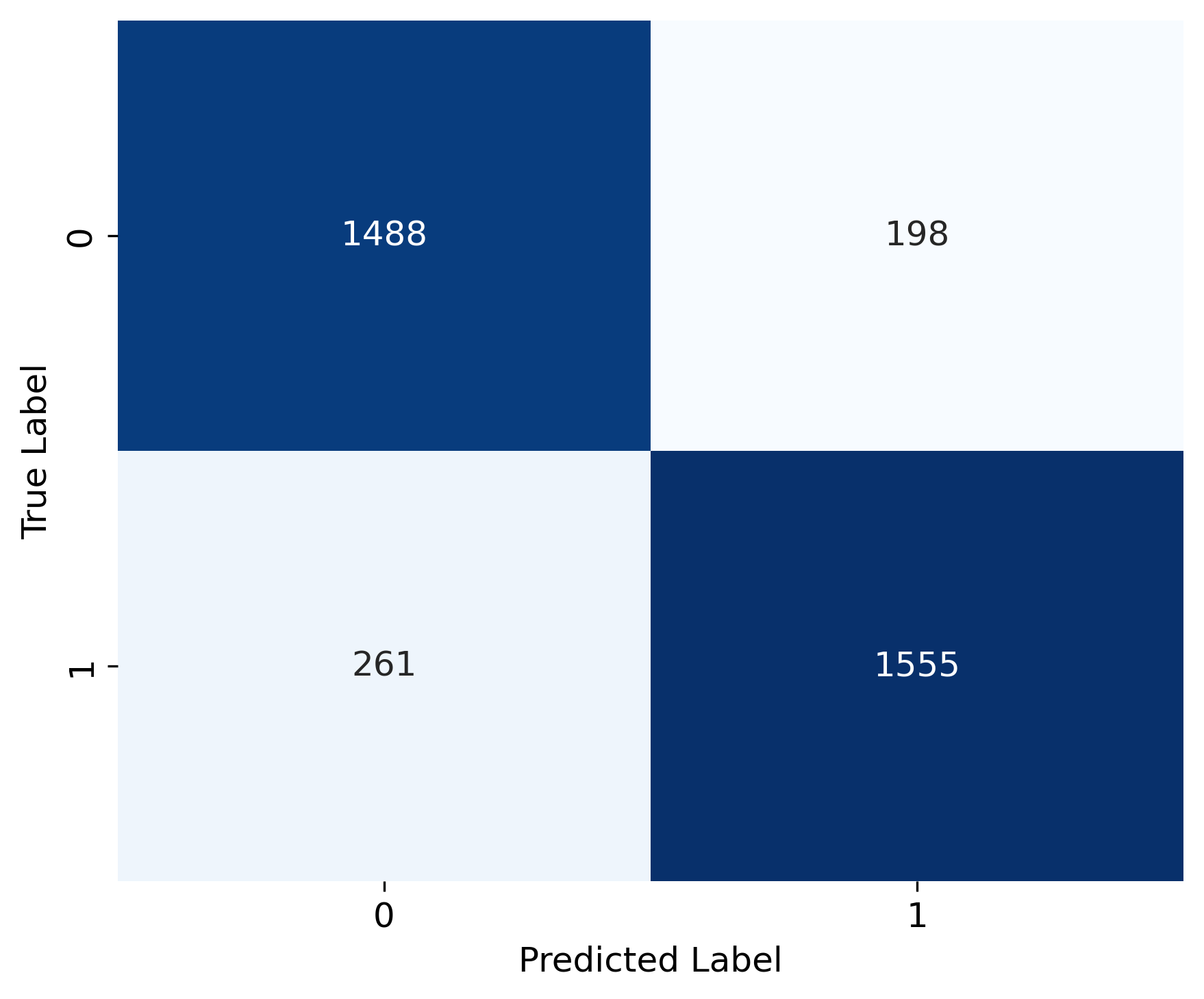} 
    \caption{Confusion Matrix for CNN Model.} 
    \label{fig:confusion_matrix_CNN} 
\end{figure}

The classification model was evaluated using trajectory-based features, with its training and validation performances depicted in the first row of Figure~\ref{fig:performance}. The loss curves (Figure~\ref{fig:performance}a) exhibit a rapid initial decline within the first few epochs, followed by gradual stabilization after $\approx$10 epochs, indicating effective learning and convergence. Similarly, the accuracy curves (Figure~\ref{fig:performance}b) for both the training and validation sets increase steadily and converge, suggesting minimal overfitting and robust generalization.

To further assess the model's generalization capability, its performance was evaluated on an unseen test set. The second row of Figure~\ref{fig:performance} presents the corresponding results. The precision-recall (PR) curve (Figure~\ref{fig:performance}c) demonstrates that the classifier maintains high precision across a wide range of recall values, with an average precision (AP) of 0.94. Similarly, the receiver operating characteristic (ROC) curve (Figure~\ref{fig:performance}d) shows a high true positive rate across varying false positive rates, with an area under the ROC curve (AUC-ROC) of 0.94, confirming the model’s strong discriminative capability.

At the final epoch, the training loss was 0.3729 and the training accuracy reached 0.8722, while the validation loss and accuracy were 0.3565 and 0.8866, respectively. The close alignment between the training and validation metrics indicates that the model generalized effectively without overfitting, consistent with similar findings in recent studies \citep{bacu2023assessment}.

To further evaluate generalization, classification metrics were assessed on the test set. The model achieved a test loss of 0.3874 and a test accuracy of 0.8689. Although slightly lower than the validation results, this minor decrease is expected, as the test set may contain more complex or ambiguous instances not fully represented in the training data \citep{roelofs2019measuring, brownlee2020identify}. Nevertheless, the model maintains strong predictive ability on unseen data, as further evidenced by the confusion matrix shown in Figure~\ref{fig:confusion_matrix_CNN}.

While \cite{rosu2021asteroid} reported a recall of 94\% for detecting asteroids in telescope imagery using a CNN model, their classification task was fundamentally different in nature. Their model was trained on static visual data to determine the presence or absence of asteroids in individual images, where pixel-level cues offer distinct patterns for detection. In contrast, our approach focuses on predicting the dynamical outcome, specifically the long-term ejection of asteroids using trajectory-based features spanning 1 Myr. This introduces significantly more complexity, driven by chaotic perturbations and nonlinear orbital evolution. Despite these challenges, our CNN model achieved a recall of approximately 88\%, highlighting its strong ability to generalize in a high-dimensional, time-dependent setting. Similarly, \cite{liu2021stability} demonstrated that a probabilistic neural network (PNN) could predict the stability of main-belt asteroids using orbital elements as inputs, achieving an overall accuracy of 87\%, which improved further when the network was trained separately for regions near mean-motion resonances. These findings demonstrate that deep learning architectures can effectively capture underlying dynamical behaviours, reinforcing their relevance for long-term orbital classification problems.

The overall evaluation suggests that the model achieves consistent and reliable results across the training, validation, and test sets. The F1 score for the training set was recorded as 0.8845, reflecting strong precision and recall. The validation F1 score remained high at 0.8859, further demonstrating effective generalization. On the test set, the F1 score was $\approx$ 0.8664, which, while slightly lower, still confirms the model’s robust performance \citep{goodfellow2016deep}. This marginal drop does not indicate overfitting but rather reflects the inherent complexity of the test data. Despite these challenges, the model consistently delivers high classification performance across all datasets.

\subsection{Comparison of the Two Approaches}

\subsubsection{Feature Inputs}

For traditional ML models, the input features comprise the initial orbital parameters of the asteroids, which serve as the foundation for numerical integration. These features include the semi-major axis, eccentricity, perigee, and other relevant parameters, which are directly provided to the model as numerical values \citep{srivastava2025classification}. 

In contrast, the CNN-based approach employs RPs as input features, derived from the concatenated time series values of the semi-major axis and eccentricity over $2 \times 10^{5}$ years. Unlike traditional ML models, where raw numerical data is used, the CNN model does not receive direct numerical inputs. Instead, RPs are generated to capture the dynamical evolution of the asteroids, emphasizing their temporal behaviour. The grayscale values of these RPs, resized to an input dimension of (150, 150, 1), are then fed into the model. This fundamental difference in feature representation underscores the different methodological approaches of the two classification strategies \citep{chauhan20182018}. \\

\subsubsection{Model Performance}

The top-performing traditional ML models including RF, AdaBoost, and GB achieved classification performance comparable to that of the CNN model, albeit with marginally lower accuracy \citep{carruba2022machine}. These results underscore the effectiveness of orbital elements in capturing salient features relevant to ejection prediction.

While traditional models operate on a limited set of static parameters, the CNN leverages a richer feature space extracted from RP images that encode dynamical behaviour. This expanded representation may enable the CNN to detect subtler patterns in temporal evolution. Nevertheless, the comparable performance of both approaches suggests that essential predictive information is already embedded within the initial orbital conditions. \\

\subsubsection{Generalization Capability}

Both traditional ML models and the CNN model exhibited strong generalization capabilities, achieved through extensive hyperparameter tuning and the effective implementation of regularization techniques. While the CNN model outperformed the best-performing ML models on the unseen test set in terms of accuracy, the difference in performance remained marginal. This suggests that, despite the differences in feature representations and learning paradigms, both approaches demonstrated comparable generalization abilities \citep{geron2022hands}. \\

\subsubsection{Computational Complexity}

The ML models, such as RF, AdaBoost, and GB, relied solely on numerical features and exhibited high computational efficiency in both training and inference. In contrast, the CNN model required significantly greater computational resources due to the preprocessing steps, the complexity of the DL architecture, and the necessity of GPU acceleration \citep{sze2017efficient}. 

Despite this disparity, the performance difference between the two approaches was marginal, with the CNN model showing only slight improvements. This trade-off underscores that while CNN models can capture complex patterns, traditional ML approaches remain competitive, achieving nearly equivalent accuracy with significantly lower computational costs.

\subsubsection{Physical Interpretability}

In the case of traditional ML models, the high evaluation metrics suggest that an asteroid's likelihood of ejection within a 1~Myr timeframe strongly depends on its initial orbital parameters at a given epoch. The high precision and recall values further indicate that both ejected and non-ejected asteroids are well-distinguished based on these initial conditions, highlighting the predictive power of orbital elements \citep{mako2005classification}. \\

Similarly, RPs were generated for the CNN model using the evolution of the semi-major axis and eccentricity over a 0.2~Myr period. The model then classifies long-term ejection outcomes using dynamical patterns extracted from the 0.2~Myr recurrence representations, leveraging the temporal structure of these parameters. The high evaluation scores, particularly in precision and recall, suggest that the recurrence behaviour of the semi-major axis and eccentricity plays a crucial role in determining the long-term stability and potential ejection of asteroids. This indicates that dynamical evolution patterns, rather than just static initial conditions, serve as key features for accurate ejection prediction \citep{alves2025deep}.

In summary, while the CNN model introduces a more complex architecture and richer dynamical features via RPs, traditional ML models trained on initial orbital parameters remain highly competitive. This highlights the sufficiency of orbital elements in ejection prediction and underscores the value of interpretable, computationally efficient approaches in celestial dynamics.

\subsection{Ejection Dynamics and Source-Region Signatures of Near-Earth Asteroids}

This subsection integrates temporal ejection statistics with forward and backward orbital integrations and model generalization, providing a unified, ensemble-level interpretation of ejection patterns and their dynamical context.

\subsubsection{Temporal Patterns in Asteroid Ejections}

\begin{figure}
    \centering
    \includegraphics[width=0.48\textwidth]{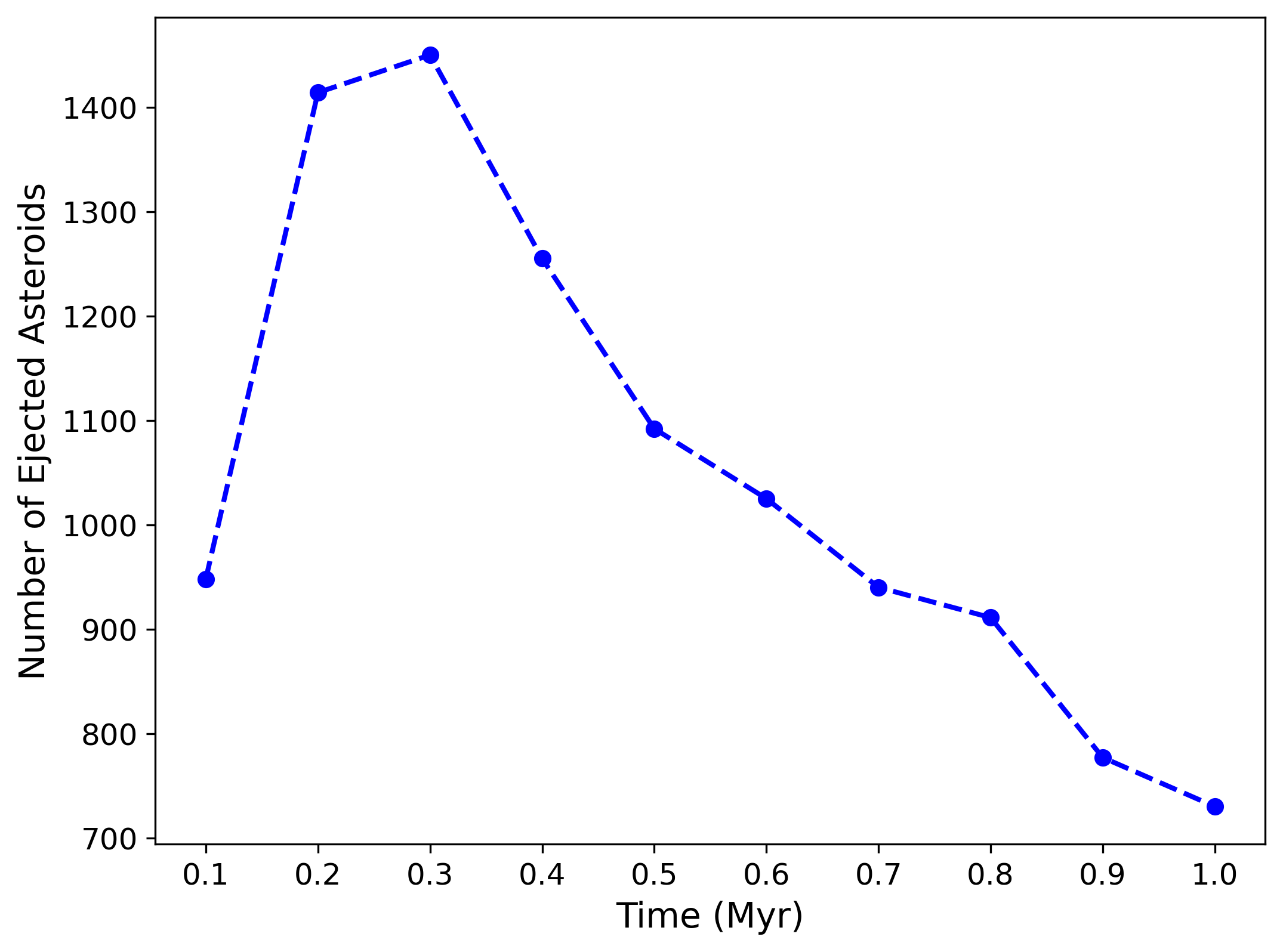}
    \caption{Number of asteroid ejections per 100,000-year interval over a one-million-year timespan.}
    \label{fig:ejection_plot}
\end{figure}

Figure~\ref{fig:ejection_plot} illustrates the number of asteroid ejections per 0.1 Myr interval across a Myr period. The x-axis represents time in Myr, while the y-axis indicates the number of asteroids ejected within each interval. The data points, connected by a dashed blue line, reveal a distinct temporal trend in ejection rates. The statistics shown here are derived from the complete integrated dataset, including both real and synthetically generated NEAs, comprising approximately 70,000 integration cases.

During the early phase (0.1-0.3~Myr), the number of ejections rises sharply, peaking at approximately 1,500 asteroids around 0.3 Myr \citep{wisdom1987chaotic}. This rapid increase suggests that a substantial fraction of NEAs were dynamically unstable from the outset, likely associated with highly chaotic or resonant regions of phase space that facilitated early removal due to gravitational perturbations \citep{gladman1997dynamical}.

Following the peak, the ejection rate declines, dropping to around 1,100 asteroids by 0.5~Myr. This decreasing trend reflects a transition from the rapid elimination of unstable asteroids to a phase where the remaining population, being comparatively more stable, required longer timescales for ejection.

Beyond 0.5~Myr, the number of ejections continues to decrease gradually. By the end of the 1~Myr period, fewer than 800 asteroids are removed per 0.1~Myr interval. This steady decline implies that the surviving asteroids reside in more stable orbital configurations and require stronger perturbations or extended timescales for ejection \citep{milani1997stable}.

Overall, we find that approximately 15\% of the NEA population is removed on Myr timescales, reflecting the prevalence of dynamically unstable orbits within the near-Earth region. The observed temporal pattern indicates that asteroid ejections are not uniformly distributed, but rather follow a characteristic progression: an early phase of rapid depletion followed by a gradual decline as the system dynamically stabilizes. These findings underscore the critical role of gravitational perturbations and chaotic evolution in shaping the long-term fate of NEAs.

\subsubsection{Ejected and Non-Ejected Asteroids of Different NEA Groups}

\begin{figure}
    \centering
    \includegraphics[width=0.48\textwidth]{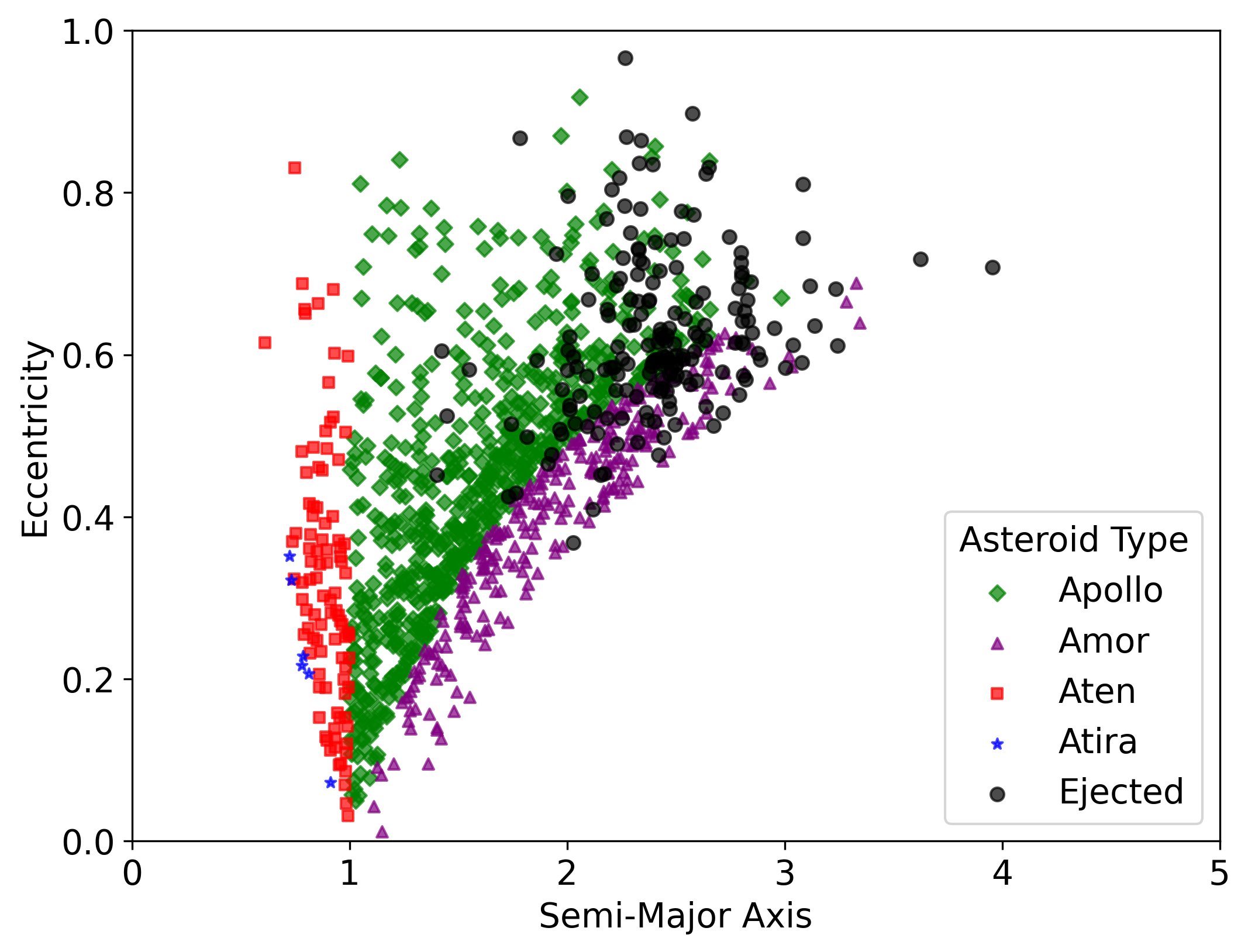}
    \caption{Distribution of semi-major axis versus eccentricity for ejected and non-ejected asteroids across different NEA subgroups.} 
    \label{fig:NEAs_ejected_non_ejected} 
\end{figure}

Figure~\ref{fig:NEAs_ejected_non_ejected} presents the semi-major axis versus eccentricity distribution for NEAs, differentiating between ejected and non-ejected asteroids. The non-ejected NEAs are categorized into four major groups Apollos, Atens, Amors, and Atiras based on their orbital characteristics \citep{bottke2000understanding}. Specifically, Apollos are depicted as green diamonds, Amors as purple triangles, Atens as red squares, and Atiras as blue stars. Ejected asteroids are represented by black circles. The plot reveals clear patterns in ejection behaviour, indicating that the majority of ejected asteroids belong to the Apollo and Amor classes.

The ejected asteroids exhibit a distinct concentration within specific semi-major axis and eccentricity ranges, highlighting the role of orbital dynamics in governing ejection likelihood \citep{liu2023orbital}. Most ejected asteroids are clustered within semi-major axes between approximately 1.5 AU and 3.5 AU and eccentricities between 0.4 and 0.8. Notably, the concentration of ejected asteroids suggests that Apollos and Amors within this semi-major axis and eccentricity range are more dynamically unstable, rendering them more susceptible to ejection. In contrast, Atens and Atiras show limited ejection activity, as evidenced by the near absence of black circles in their respective orbital domains. This pattern implies that orbital configurations associated with moderate to high eccentricities and semi-major axes greater than 1.5 AU are particularly vulnerable to gravitational perturbations and subsequent ejection. The clustering of ejected asteroids within this restricted orbital range underscores the significance of initial orbital conditions and dynamical interactions in determining the long-term stability of NEAs.

\subsubsection{Backward Integration and Dynamical Behaviour of Ejected Asteroids}

\begin{figure*}
\centering
\hspace*{-3mm}

\begin{minipage}{0.48\textwidth}
    \centering
    \includegraphics[width=\linewidth]{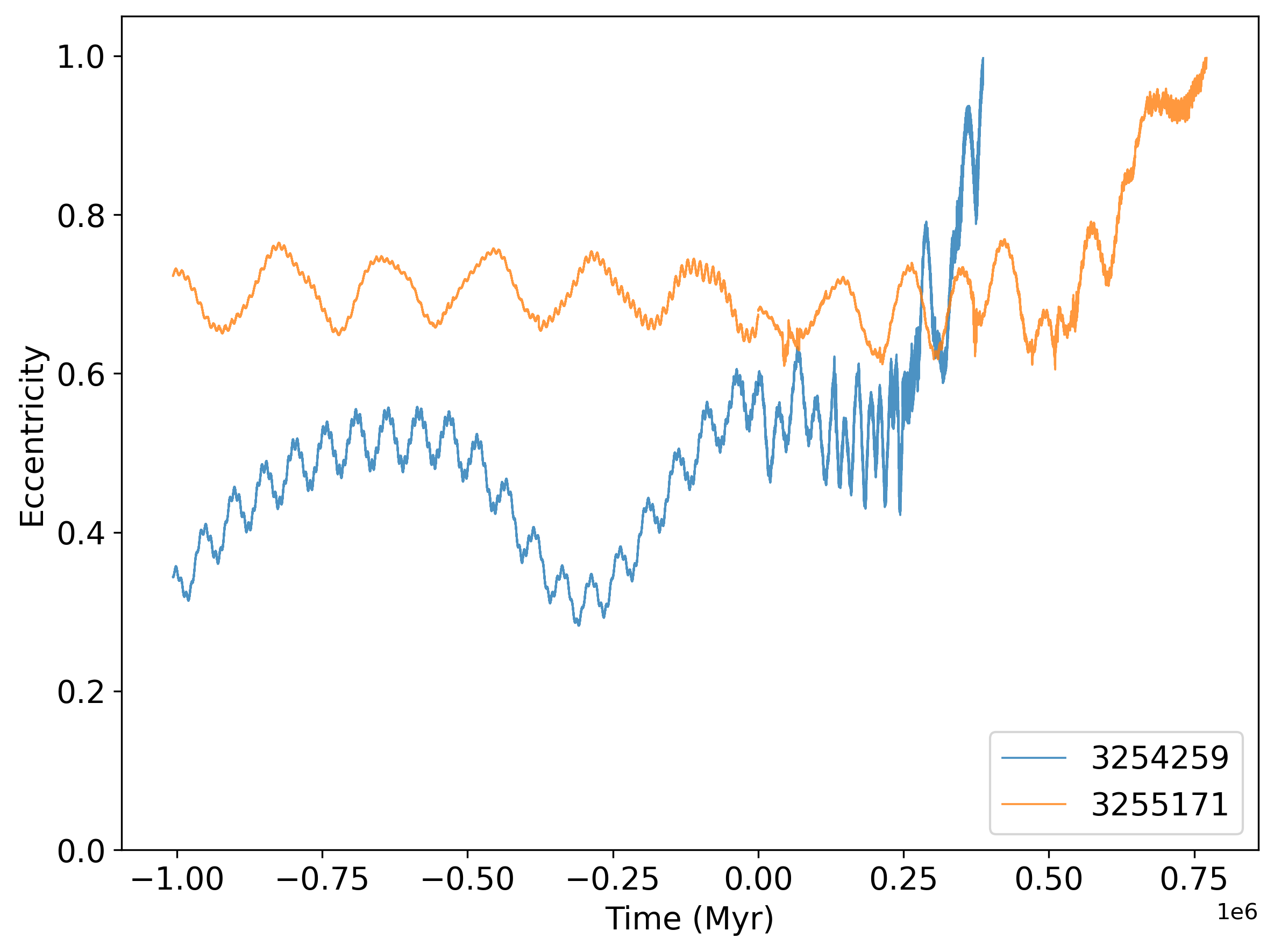}

    \vspace{0.5ex}
    (a)
\end{minipage}\hspace{0.5cm}
\begin{minipage}{0.48\textwidth}
    \centering
    \includegraphics[width=\linewidth]{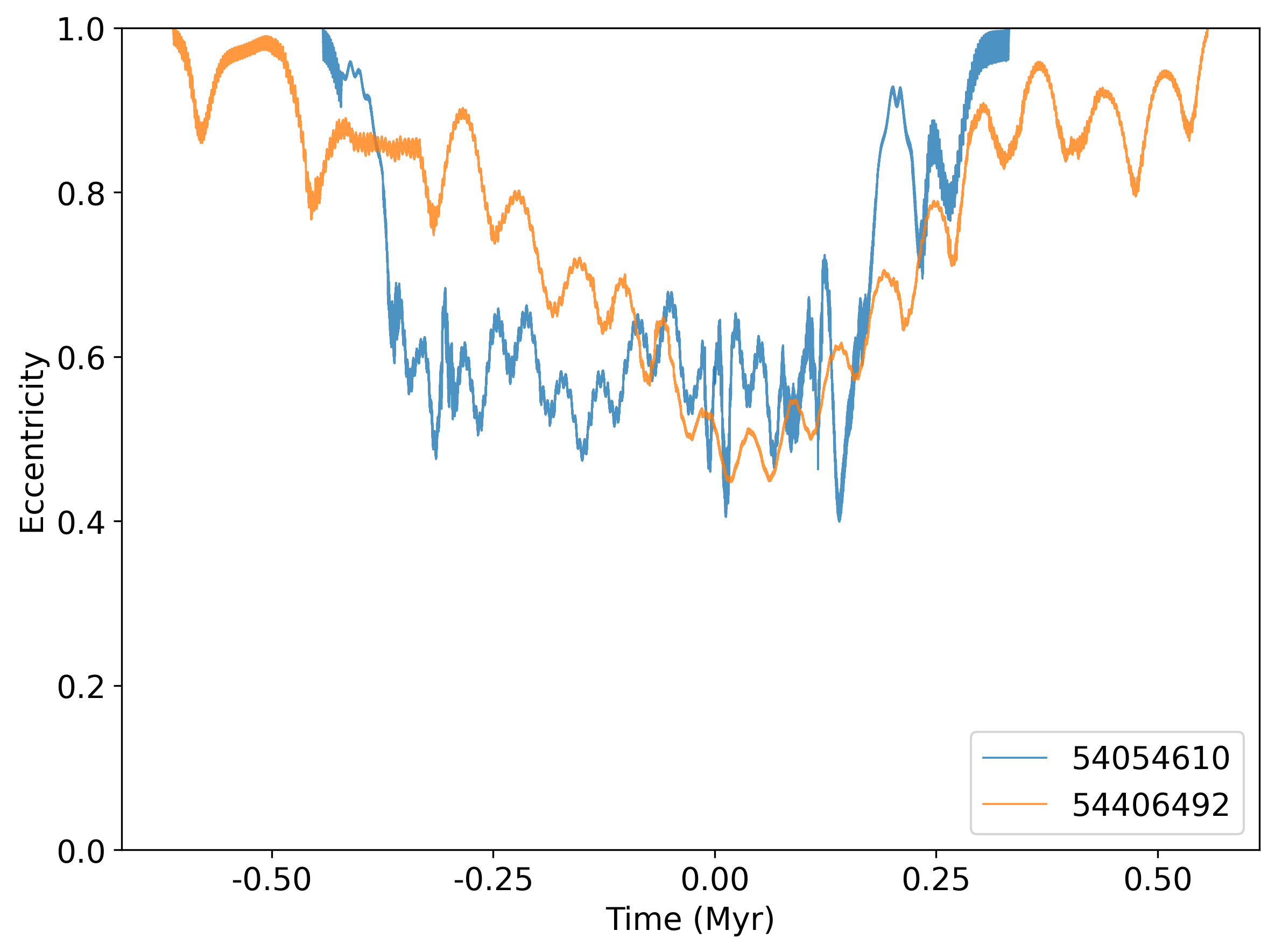}

    \vspace{0.5ex}
    (b)
\end{minipage}

\vspace{0.4cm}

\begin{minipage}{0.48\textwidth}
    \centering
    \includegraphics[width=\linewidth]{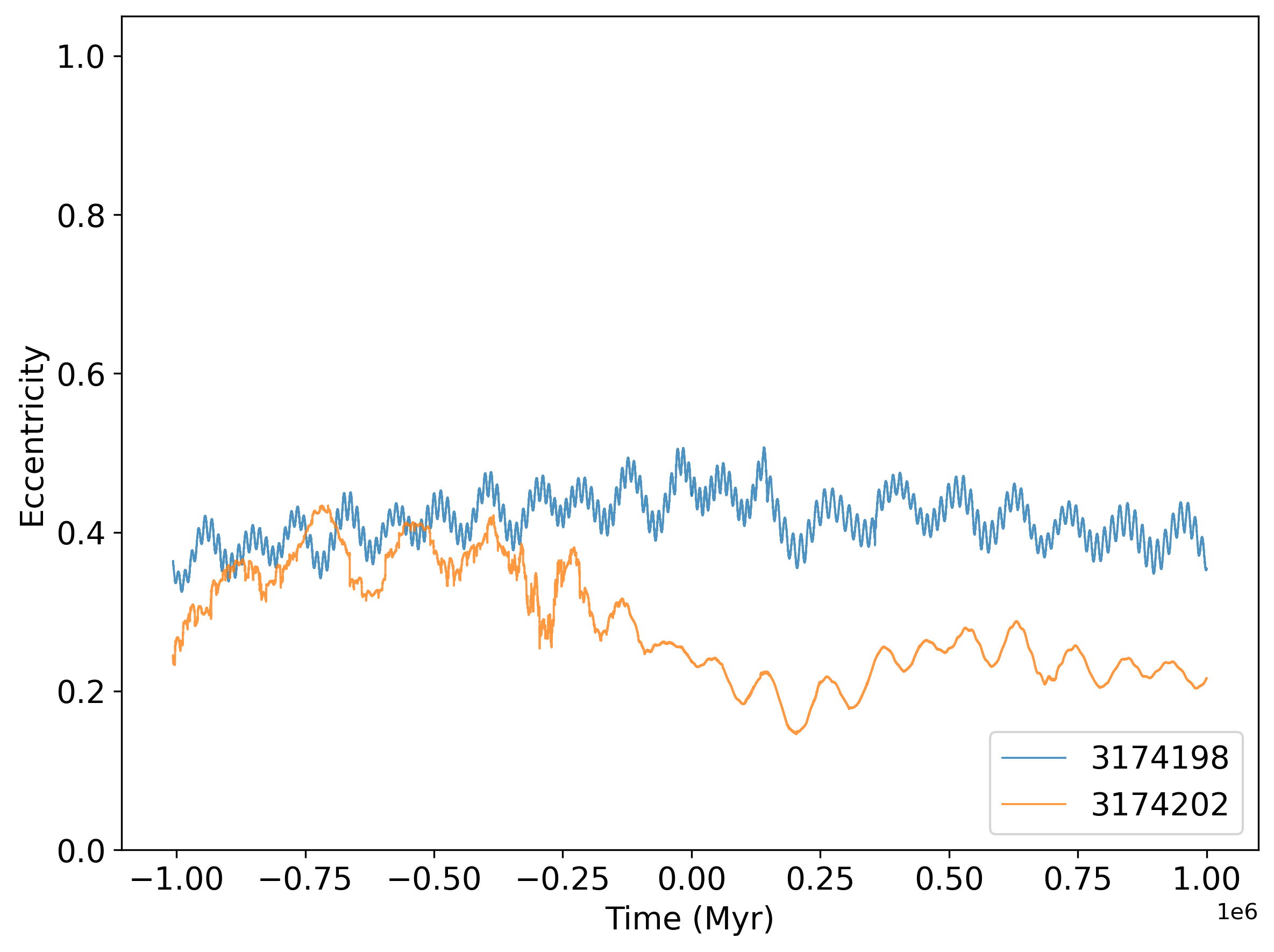}

    \vspace{0.5ex}
    (c)
\end{minipage}\hspace{0.5cm}
\begin{minipage}{0.48\textwidth}
    \centering
    \includegraphics[width=\linewidth]{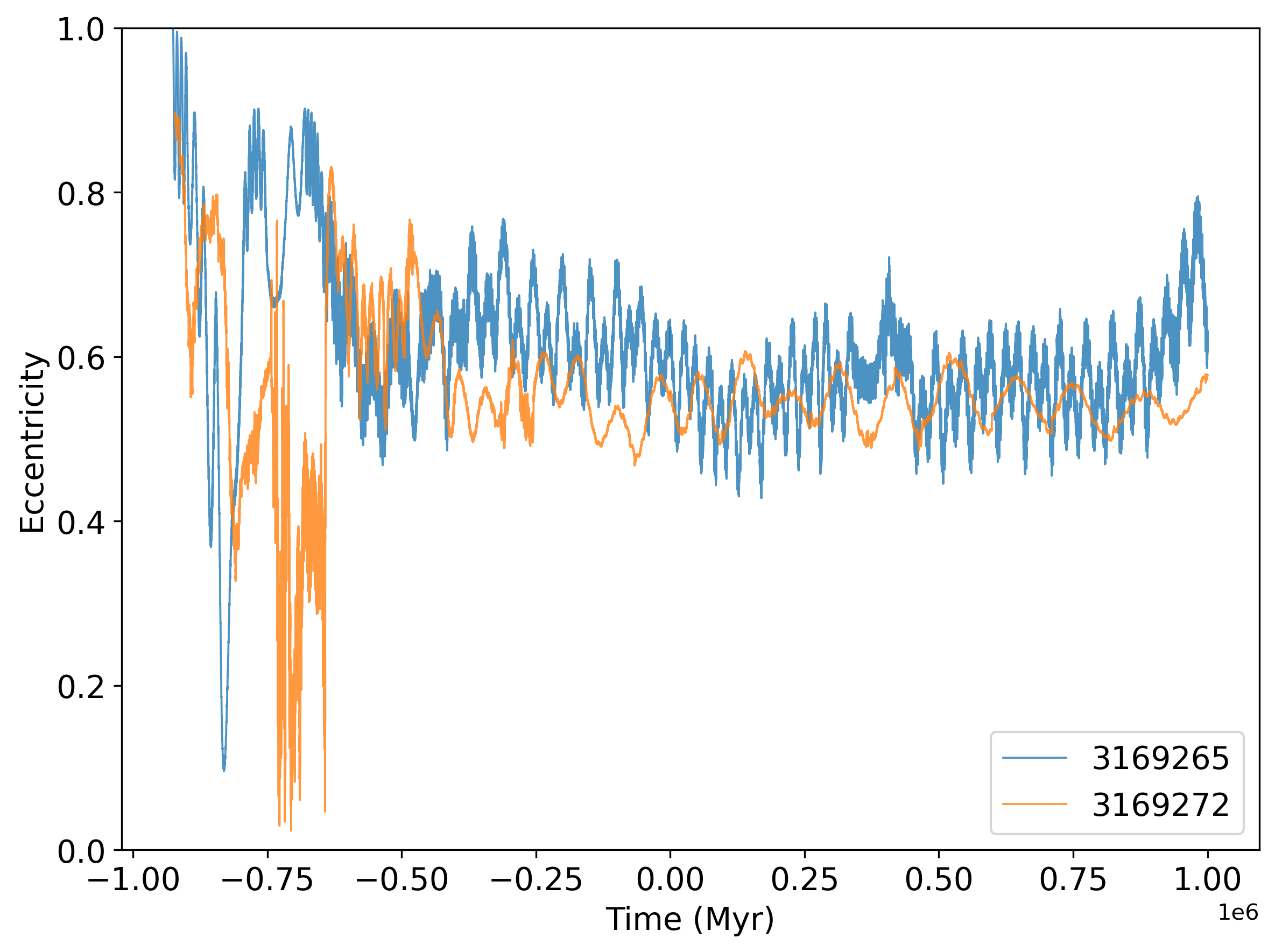}

    \vspace{0.5ex}
    (d)
\end{minipage}

\caption{Time evolution of orbital eccentricity for different ejection outcomes in forward and backward integrations.
(a) Ejection (forward): asteroids ejected in the forward integration but remaining bound in the backward integration.
(b) Ejection (forward and backward): asteroids ejected in both integration directions.
(c) No ejection: asteroids that remain bound in both forward and backward integrations.
(d) Ejection (backward): asteroids ejected in the backward integration but remaining bound in the forward integration.
Time zero corresponds to the present epoch; negative and positive times denote backward and forward propagation, respectively.}
\label{fig:ecc_vs_time}
\end{figure*}

To explore the statistical implications of backward integrations for NEA dynamical histories, we performed a backward integration of a subset over a period of 1 Myr. This analysis is used as a diagnostic tool to examine ensemble-level stability and instability patterns. The same ejection criteria as adopted for the forward integrations were applied.

The backward integration revealed that approximately 15\% of the asteroids satisfied the ejection criterion. This indicates that a non-negligible fraction of the ensemble occupies orbital configurations that are dynamically unstable when propagated backward in time, consistent with strong chaotic sensitivity. In contrast, about 85\% of the asteroids remained bound within the inner Solar System, reflecting comparatively stable configurations over the finite integration interval.

A notable result from the comparison of forward and backward integrations is the presence of noticeable overlap between objects ejected in both temporal directions. This overlap identifies a subset of asteroids residing in highly unstable regions of phase space, for which ejection occurs readily regardless of the direction of time propagation. In contrast, a complementary population of asteroids is ejected only in the forward integration while remaining bound when propagated backward, consistent with recent destabilization from previously longer-lived orbital configurations \citep{gladman2000near}.

Among the asteroids that remained bound during backward integration, a large fraction were classified as NEAs at the reference epoch, primarily belonging to the Apollo, Amor, Aten, and Atira groups. The remainder included objects dynamically associated with the Mars-crossing and main-belt populations, with smaller contributions from the outer belt and trans-Neptunian region \citep{dvorak1993survey, morbidelli2015dynamical}. These distributions emphasize the diversity of dynamical pathways feeding the NEA population and illustrate how backward integrations can be used to characterize ensemble-level instability.

\subsubsection{Performance of CNN and ML Models on Backward-Integrated Data}

To evaluate model generalization, we applied the previously trained CNN model developed using forward integration data to classify asteroid ejection outcomes in the backward direction. The same preprocessing pipeline was maintained. While the CNN preserved high recall, its precision declined significantly, marking a departure from its strong forward performance. Such behaviour is consistent with known limitations of deep learning models when confronted with out-of-distribution inputs or shifted dynamical regimes \citep{quinonero2022dataset, amodei2016concrete, taori2020measuring}. This drop in accuracy likely results from a distribution shift in the input features during backward integration, which the model was not exposed to during training.

In contrast, the top-performing traditional ML models RF, AdaBoost, and GB achieved both high recall and good precision when applied to the backward dataset. Since these models rely solely on the initial orbital conditions, which remain unchanged across integration directions, they demonstrated greater resilience to distributional changes. Their robust performance reinforces the predictive power of static orbital features.

\subsubsection{Dynamical Case Studies of Eccentricity Evolution}

To better understand the mechanisms driving asteroid stability and ejection, Figure~\ref{fig:ecc_vs_time} presents eccentricity-time profiles for representative NEAs, integrated over a 2~Myr interval ($\pm1$~Myr). Time zero corresponds to the present epoch; negative values denote backward propagation, and positive values denote forward propagation.

(i) \textbf{Ejected Only in Forward Direction}:  
Figure~\ref{fig:ecc_vs_time}a shows the eccentricity evolution of asteroids 2004 SY4 (3254259) and 2004 ST26 (3255171). During backward integration, both exhibit relatively stable behaviour: 3254259 oscillates between $e \approx 0.3$ and 0.6 with low amplitude, while 3255171 maintains $e \approx 0.7$ with minimal variation. These trends are indicative of modest gravitational perturbations, possibly regulated by weak secular resonances (e.g., $\nu_5$, $\nu_6$) \citep{williams1981positions}. However, forward integration reveals strong dynamical excitation beginning near +0.25 Myr, with eccentricities rising rapidly toward $e \approx 1.0$, consistent with resonance crossings or close planetary encounters leading to orbital instability and eventual ejection. This asymmetry underscores the sensitivity of NEA orbits to slow diffusion and cumulative perturbations over Myr timescales.

(ii) \textbf{Ejected in Both Directions}:
In Figure~\ref{fig:ecc_vs_time}b, asteroids 54054610 (2020 SL4) and 54406492 (2023 WG) exhibit strongly chaotic eccentricity behaviour. Backward integrations for both objects reveal a persistent growth of eccentricity toward near-parabolic values ($e \rightarrow 1$) that terminates in ejection, a behaviour consistent with dynamical pathways often associated with objects originating exterior to the inner Solar System and undergoing transient residence in the near-Earth region. The monotonic eccentricity increase during backward propagation is indicative of dynamical behaviour associated with transitions between bound and unbound motion in chaotic regimes. Forward integrations display comparable large-amplitude, irregular oscillations and a rapid march toward escape, so that instability is apparent in both temporal directions. The symmetry of this unstable behaviour indicates short dynamical lifetimes for these bodies and supports their interpretation as short-lived, dynamically unstable objects subject to repeated resonant forcing and scattering by planets and mean-motion resonances \citep{migliorini1995interlopers}.

(iii) \textbf{Stable in Both Directions}:  
Figure~\ref{fig:ecc_vs_time}c presents the eccentricity evolution of 2004 BB75 (3174198) and 2004 BE86 (3174202), which remain gravitationally bound throughout the full interval. Asteroid 3174198 displays consistent oscillations between $e \approx 0.35$ and 0.50, while 3174202 gradually declines from $e \approx 0.25$ to 0.20. The low-amplitude, periodic behaviour in both cases indicates residence within dynamically quiescent regions, likely protected by weak mean-motion or secular resonances. The lack of strong excursions or chaotic transitions suggests long-term orbital stability, making these objects typical of NEAs with extended dynamical lifetimes \citep{milani1997stable}.

(iv) \textbf{Ejected Only in Backward Direction}:  
Figure~\ref{fig:ecc_vs_time}d shows asteroids 2003 WC158 (3169265) and 2003 XV (3169272), both of which experience significant eccentricity growth during backward integration, reaching $e \approx 1.0$ before being ejected \citep{bottke2006yarkovsky}. In forward time, however, their orbits remain relatively stable with eccentricities between 0.5 and 0.7. This asymmetric behaviour is consistent with scenarios in which objects evolve from highly chaotic configurations toward comparatively more stable states, suggesting dynamical pathways compatible with capture-like behaviour from more distant regions of phase space. Such behaviour may be associated with gravitational scattering processes or temporary residence in stabilizing resonances. Over time, mechanisms such as phase space diffusion and resonance sticking could have contributed to their gradual stabilization.

\vspace{1ex}

These case studies emphasize the wide range of dynamical behaviours within the NEA population. The comparison of forward and backward integrations reveals asymmetric evolutionary pathways, with some objects exhibiting persistent instability while others evolve from chaotic origins into stable configurations. Eccentricity evolution serves as a key diagnostic for identifying past capture, future ejection, and the influence of planetary perturbations across Solar System timescales.

\section{Discussion and Conclusions}
\label{sec:discussion_conclusions}

This work examined whether data-driven approaches can serve as efficient proxies for computationally expensive long-term $N$-body integrations in identifying NEAs that are removed from the Solar System within $1\,$Myr. We compared tree-ensemble classifiers trained on instantaneous orbital elements with a CNN trained on RPs constructed from concatenated semi-major axis and eccentricity time series over a $0.2\,$Myr window.

Both methodological families produce strong and consistent results. Tree-ensemble methods (Random Forest, AdaBoost, Gradient Boosting) trained on the initial ephemerides deliver rapid, interpretable predictions with balanced performance (test accuracy $\simeq0.86$, recall $\simeq0.91$, $F_{1}$ score $\simeq0.87$). Feature-importance analysis indicates that semi-major axis and eccentricity provide the majority of the discriminative signal (roughly $72\%$ and $13\%$, respectively). The CNN operating on $150\times150$ grayscale RPs achieves comparable test performance (accuracy $\simeq0.87$, AP$\simeq0.94$, AUC$\simeq0.94$, $F_{1}$ score $\simeq0.87$), demonstrating that short-term dynamical morphology encodes signatures associated with long-term instability. These results indicate that much of the information relevant to 1-Myr ejection is captured either directly by initial orbital geometry or indirectly by short-term recurrence structure.

Dynamically, the ensemble of ejected objects is concentrated at larger semi-major axes ($a\sim1.5$--$3.5\,$AU) and elevated eccentricities ($e\sim0.4$--$0.8$), consistent with resonance-driven diffusion and enhanced likelihood of close planetary encounters. The temporal distribution of ejections is non-uniform: ejection rates peak early (approximately $0.2$–$0.4\,$Myr) and decline thereafter, producing an overall ejection fraction of order $\sim15\%$ on Myr timescales. This is consistent with the dynamically chaotic nature of NEAs, whose lifetimes are typically much shorter than Solar System evolutionary timescales. This concentration of early removals explains, in part, why short-term dynamical indicators are effective predictors for a large subset of unstable objects. Backward integrations yield a comparable ejection fraction over $1\,$Myr, while the surviving population is dominated by objects belonging to established NEA dynamical groups (Apollo, Amor, Aten, and Atira), with additional contributions dynamically associated with Mars-crossing and main-belt populations and smaller fractions linked to inner-belt, outer-belt, and trans-Neptunian source regions.

Several important caveats are as follows. First, the CNN is trained on a fixed 0.2~Myr window while the labels refer to outcomes at 1~Myr; consequently, slow, late-onset instabilities driven by gradual resonance diffusion or delayed close encounters are less likely to be anticipated by short-window recurrence morphology, producing a portion of false negatives. Second, label noise arising from finite observational uncertainty is non-negligible: Monte-Carlo cloning (50 realizations per object) shows that a notable minority of nominal orbits produce divergent macroscopic outcomes across clones, implying intrinsic epistemic uncertainty for some training labels. Third, although the dataset augmentation procedures were designed to preserve the statistical and dynamical properties of the underlying NEA population, augmentation based on perturbed orbital elements and temporal segmentation necessarily represents an approximate sampling of the local phase space. Such procedures are standard in time-series learning and improve model generalization, but the resulting training distribution may still differ subtly from the distribution encountered in independent dynamical realizations. Finally, the adopted numerical conventions and modelling choices, operational ejection threshold (100~AU or hyperbolic escape), disabling of collisional removal, and simplified parameterizations of non-gravitational forces, affect absolute ejection counts and should be borne in mind when interpreting quantitative fractions.

Taken together, the results support a practical hybrid strategy: employ tree-ensemble classifiers on initial orbital parameters for large-scale, low-cost screening and prioritization, and reserve trajectory-based (RP/CNN or other time-aware) analyses for targeted follow-up of ambiguous or scientifically high-value objects. While these data-driven tools are not substitutes for high-accuracy long-term integrations when precise dynamical evolution must be resolved, they provide effective, scalable diagnostics for triaging objects and guiding allocation of computational resources for detailed dynamical study. Future work could refine probabilistic labelling using clone ensembles, explore variable-length and attention-based time-series architectures to better capture slow diffusion, and quantify sensitivity to the numerical modelling choices noted above.

\section*{Acknowledgements}

We sincerely thank the Inter-University Centre for Astronomy and Astrophysics (IUCAA) in Pune for providing access to their research facilities through the Visiting Associateship program. We also express our gratitude to the Indian Space Research Organisation (ISRO), Department of Space, Government of India, for their financial support under the RESPOND program (ISRO/RES/3/1034/24-25).

\section*{Data and Code Availability}

We acknowledge the use of data from NASA JPL Solar System Dynamics (Downloaded 2024, November 21; Small-Body Database Query; \url{https://ssd.jpl.nasa.gov/}). The raw asteroid orbital data used in this study are publicly available from this source. The processed dataset used for training and evaluation has been deposited on Zenodo at \url{https://doi.org/10.5281/zenodo.19554245}. The Zenodo record also includes the trained CNN model weights (\texttt{best\_model.h5}) to enable direct reproducibility of the deep-learning results.

The code used for training and evaluating the machine-learning and deep-learning models is available in a public GitHub repository at \url{https://github.com/chetan-hub-02/asteroid-dynamics-ml-dl}. The repository includes the training and evaluation scripts, model weights, and implementation details required to inspect and reproduce the analysis.

\begin{contribution}

All authors contributed equally.


\end{contribution}

%



\bibliography{chetan}{}

@inproceedings{bottke1993collision,
  title={Collision lifetimes and impact statistics of near-Earth asteroids},
  author={Bottke Jr, WF and Nolan, Michael C and Greenberg, Richard},
  booktitle={In Lunar and Planetary Inst., Twenty-fourth Lunar and Planetary Science Conference. Part 1: AF p 159-160 (SEE N94-12015 01-91)},
  volume={24},
  year={1993}
}

@article{bottke2002debiased,
	title={Debiased orbital and absolute magnitude distribution of the near-Earth objects},
	author={Bottke Jr, William F and Morbidelli, Alessandro and Jedicke, Robert and Petit, Jean-Marc and Levison, Harold F and Michel, Patrick and Metcalfe, Travis S},
	journal={Icarus},
	volume={156},
	number={2},
	pages={399--433},
	year={2002},
	publisher={Elsevier}
}

@article{perna2013near,
  title={The near-Earth objects and their potential threat to our planet},
  author={Perna, D and Barucci, MA and Fulchignoni, M},
  journal={The Astronomy and Astrophysics Review},
  volume={21},
  number={1},
  pages={65},
  year={2013},
  publisher={Springer}
}

@article{carruba2025invisible,
  title={The invisible threat-Assessing the collisional hazard posed by undiscovered Venus co-orbital asteroids},
  author={Carruba, Valerio and Sfair, Rafael and Araujo, Rosana AN and Winter, Othon C and Mour{\~a}o, Daniela C and Di Ruzza, Sara and Aljbaae, Safwan and Carit{\'a}, Gabriel and Domingos, Rita C and Alves, AA},
  journal={Astronomy \& Astrophysics},
  volume={699},
  pages={A86},
  year={2025},
  publisher={EDP Sciences}
}

@article{bulirsch1966numerical,
  title={Numerical treatment of ordinary differential equations by extrapolation methods},
  author={Bulirsch, Roland and Stoer, Josef},
  journal={Numerische Mathematik},
  volume={8},
  number={1},
  pages={1--13},
  year={1966},
  publisher={Springer}
}

@article{rein2015ias15,
  title={IAS15: a fast, adaptive, high-order integrator for gravitational dynamics, accurate to machine precision over a billion orbits},
  author={Rein, Hanno and Spiegel, David S},
  journal={Monthly Notices of the Royal Astronomical Society},
  volume={446},
  number={2},
  pages={1424--1437},
  year={2015},
  publisher={Oxford University Press}
}

@article{carita2024image,
  title={Image classification of retrograde resonance in the planar circular restricted three-body problem},
  author={Carit{\'a}, GA and Aljbaae, Safwan and Morais, Maria Helena Moreira and Signor, Alan Cefali and Carruba, Valerio and Prado, Antonio FBA and Hussmann, Hauke},
  journal={Celestial Mechanics and Dynamical Astronomy},
  volume={136},
  number={2},
  pages={10},
  year={2024},
  publisher={Springer}
}

@article{smirnov2026implementation,
  title={Implementation of secular resonance support in the open-source python package “resonances”},
  author={Smirnov, Evgeny A},
  journal={Astronomy and Computing},
  volume={54},
  pages={101022},
  year={2026},
  publisher={Elsevier}
}

@article{carruba2025vision,
  title={Vision transformers for identifying asteroids interacting with secular resonances},
  author={Carruba, Valerio and Aljbaae, Safwan and Smirnov, Evgeny and Carit{\'a}, Gabriel},
  journal={Icarus},
  volume={425},
  pages={116346},
  year={2025},
  publisher={Elsevier}
}

@article{morbidelli2002modern,
	title={Modern integrations of solar system dynamics},
	author={Morbidelli, Alessandro},
	journal={Annual Review of Earth and Planetary Sciences},
	volume={30},
	number={1},
	pages={89--112},
	year={2002},
	publisher={Annual Reviews 4139 El Camino Way, PO Box 10139, Palo Alto, CA 94303-0139, USA}
}

@article{wang2020systematic,
	title={A systematic review of machine learning models for predicting outcomes of stroke with structured data},
	author={Wang, Wenjuan and Kiik, Martin and Peek, Niels and Curcin, Vasa and Marshall, Iain J and Rudd, Anthony G and Wang, Yanzhong and Douiri, Abdel and Wolfe, Charles D and Bray, Benjamin},
	journal={PloS one},
	volume={15},
	number={6},
	pages={e0234722},
	year={2020},
	publisher={Public Library of Science San Francisco, CA USA}
}

@article{levison2000symplectically,
	title={Symplectically Integrating Close Encounters withthe Sun},
	author={Levison, Harold F and Duncan, Martin J},
	journal={The Astronomical Journal},
	volume={120},
	number={4},
	pages={2117},
	year={2000},
	publisher={IOP Publishing}
}

@article{ismail2019deep,
	title={Deep learning for time series classification: a review},
	author={Ismail Fawaz, Hassan and Forestier, Germain and Weber, Jonathan and Idoumghar, Lhassane and Muller, Pierre-Alain},
	journal={Data mining and knowledge discovery},
	volume={33},
	number={4},
	pages={917--963},
	year={2019},
	publisher={Springer}
}

@article{chen2014big,
	title={Big data deep learning: challenges and perspectives},
	author={Chen, Xue-Wen and Lin, Xiaotong},
	journal={IEEE access},
	volume={2},
	pages={514--525},
	year={2014},
	publisher={Ieee}
}

@article{choudhary2025exoplanet,
  title={Exoplanet classification through vision transformers with temporal image analysis},
  author={Choudhary, Anupma and Bandari, Sohith and Kushvah, BS and Swastik, C},
  journal={The Astronomical Journal},
  volume={170},
  number={2},
  pages={120},
  year={2025},
  publisher={The American Astronomical Society}
}

@article{robinson2009recurrences,
	title={Recurrences determine the dynamics},
	author={Robinson, Geoffrey and Thiel, Marco},
	journal={Chaos: An Interdisciplinary Journal of Nonlinear Science},
	volume={19},
	number={2},
	year={2009},
	publisher={AIP Publishing}
}

@article{thiel2004estimation,
	title={Estimation of dynamical invariants without embedding by recurrence plots},
	author={Thiel, Marco and Romano, M Carmen and Read, PL and Kurths, J},
	journal={Chaos: An Interdisciplinary Journal of Nonlinear Science},
	volume={14},
	number={2},
	pages={234--243},
	year={2004},
	publisher={American Institute of Physics}
}

@article{webber2005recurrence,
	title={Recurrence quantification analysis of nonlinear dynamical systems},
	author={Webber Jr, Charles L and Zbilut, Joseph P},
	journal={Tutorials in contemporary nonlinear methods for the behavioral sciences},
	volume={94},
	number={2005},
	pages={26--94},
	year={2005}
}

@inproceedings{hatami2018classification,
	title={Classification of time-series images using deep convolutional neural networks},
	author={Hatami, Nima and Gavet, Yann and Debayle, Johan},
	booktitle={Tenth international conference on machine vision (ICMV 2017)},
	volume={10696},
	pages={242--249},
	year={2018},
	organization={SPIE}
}

@article{gladman1997dynamical,
	title={Dynamical lifetimes of objects injected into asteroid belt resonances},
	author={Gladman, Brett J and Migliorini, Fabbio and Morbidelli, Alessandro and Zappala, Vincenzo and Michel, Patrick and Cellino, Alberto and Froeschle, Christiane and Levison, Harold F and Bailey, Mark and Duncan, Martin},
	journal={Science},
	volume={277},
	number={5323},
	pages={197--201},
	year={1997},
	publisher={American Association for the Advancement of Science}
}

@article{nesvorny2002regular,
	title={Regular and Chaotic Dynamics in the Mean-Motion Resonances: Implications for the Structure and},
	author={Nesvorn{\`y}, D},
	journal={Asteroids III},
	pages={379},
	year={2002},
	publisher={University of Arizona Press}
}

@article{morbidelli1999numerous,
	title={Numerous weak resonances drive asteroids toward terrestrial planets orbits},
	author={Morbidelli, A and Nesvorn{\`y}, D},
	journal={Icarus},
	volume={139},
	number={2},
	pages={295--308},
	year={1999},
	publisher={Elsevier}
}

@article{bora2024temporal,
  title={Temporal trends in asteroid behaviour: a machine learning and N-body integration approach},
  author={Bora, Chetan Abhijnanam and Kushvah, Badam Singh and Mouli, Gunda Chandra and Yousuf, Saleem},
  journal={Monthly Notices of the Royal Astronomical Society},
  volume={534},
  number={1},
  pages={415--430},
  year={2024},
  publisher={Oxford University Press}
}

@inproceedings{sharma2023asteroid,
  title={Asteroid Hazard Prediction Using Machine Learning: A Comparative Analysis of Different Algorithms},
  author={Sharma, Tushar and Sharma, Shamneesh and Sharma, Ajay and Kumar, Aman and Malik, Arun and Sharma, Abhimanyu},
  booktitle={2023 3rd International Conference on Advancement in Electronics \& Communication Engineering (AECE)},
  pages={673--677},
  year={2023},
  organization={IEEE}
}

@article{malakouti2023machine,
  title={Machine learning techniques for classifying dangerous asteroids},
  author={Malakouti, Seyed Matin and Menhaj, Mohammad Bagher and Suratgar, Amir Abolfazl},
  journal={MethodsX},
  volume={11},
  pages={102337},
  year={2023},
  publisher={Elsevier}
}

@article{vujicic2020classification,
  title={Classification of asteroid families with artificial neural networks},
  author={Vujicic, D and Pavlovic, R and Milo{\v{s}}evic, D and Dor-devic, B and Ran-dic, S and Stojic, D},
  journal={Serb. Astron. J},
  volume={201},
  pages={39--48},
  year={2020}
}

@article{smirnov2017identification,
  title={Identification of asteroids trapped inside three-body mean motion resonances: a machine-learning approach},
  author={Smirnov, Evgeny A and Markov, Alexey B},
  journal={Monthly Notices of the Royal Astronomical Society},
  volume={469},
  number={2},
  pages={2024--2031},
  year={2017},
  publisher={Oxford University Press}
}

@inproceedings{vignesh2024asteroid,
	title={Asteroid Family Classification using Boosting Algorithms},
	author={Vignesh, A Sai and Meena, T and Pravardhitha, M Sai and Kennanya, M Sam},
	booktitle={2024 Asia Pacific Conference on Innovation in Technology (APCIT)},
	pages={1--6},
	year={2024},
	organization={IEEE}
}

@inproceedings{sajid2024machine,
	title={Machine Learning and Ensemble Models for Hazardous Asteroids Prediction},
	author={Sajid, Sadiya and Das, Shreyashi and Kokatnoor, Sujatha Arun},
	booktitle={International Conference on Data Science, Computation and Security},
	pages={145--155},
	year={2024},
	organization={Springer}
}

@article{izza2020explaining,
	title={On explaining decision trees},
	author={Izza, Yacine and Ignatiev, Alexey and Marques-Silva, Joao},
	journal={arXiv preprint arXiv:2010.11034},
	year={2020}
}

@article{chandra2025enhanced,
	title={Enhanced Predictive Modeling for Hazardous Near-Earth Object Detection: A Comparative Analysis of Advanced Resampling Strategies and Machine Learning Algorithms in Planetary Risk Assessment},
	author={Chandra, Sunkalp},
	journal={arXiv preprint arXiv:2508.15106},
	year={2025}
}

@inproceedings{almousa2025asteroid,
	title={Asteroid Orbit Classification with Machine Learning: A Data-Driven Approach},
	author={Almousa, Leen and Hamdan, Lojain and Hammouri, Tala and Abdel-Nabi, Heba},
	booktitle={2025 International Conference on New Trends in Computing Sciences (ICTCS)},
	pages={156--163},
	year={2025},
	organization={IEEE}
}

@article{carruba2021artificial,
  title={Artificial neural network classification of asteroids in the M1: 2 mean-motion resonance with Mars},
  author={Carruba, Valerio and Aljbaae, Safwan and Domingos, Rita C and Barletta, William},
  journal={Monthly Notices of the Royal Astronomical Society},
  volume={504},
  number={1},
  pages={692--700},
  year={2021},
  publisher={Oxford University Press}
}

@article{carruba2024digitally,
  title={Digitally filtered resonant arguments for deep learning classification of asteroids in secular resonances},
  author={Carruba, V and Aljbaae, S and Domingos, RC and Carit{\'a}, G and Alves, A and Delfino, EMDS},
  journal={Monthly Notices of the Royal Astronomical Society},
  volume={531},
  number={4},
  pages={4432--4443},
  year={2024},
  publisher={Oxford University Press}
}

@article{carruba2023deep,
  title={Deep learning classification of asteroids in g-type secular resonances},
  author={Carruba, Valerio and Aljbaae, Safwan and Knezevic, Zoran and Mahlke, Max and Masiero, Joseph and Roig, Fernando and Domingos, Rita and Huaman, Mariela and Alves, Abreu{\c{c}}on and Martins, Bruno and others},
  year={2023},
  note={Manuscript under review}
}

@article{carruba2023imbalanced,
  title={Imbalanced classification applied to asteroid resonant dynamics},
  author={Carruba, V and Aljbaae, S and Carit{\'a}, G and Louren{\c{c}}o, MVF and Martins, BS and Alves, AA},
  journal={Frontiers in Astronomy and Space Sciences},
  volume={10},
  pages={1196223},
  year={2023},
  publisher={Frontiers Media SA}
}

@article{levison1994long,
  title={The long-term dynamical behavior of short-period comets},
  author={Levison, Harold F and Duncan, Martin J},
  journal={Icarus},
  volume={108},
  number={1},
  pages={18--36},
  year={1994},
  publisher={Elsevier}
}

@article{grimm2014genga,
  title={The GENGA code: gravitational encounters in N-body simulations with GPU acceleration},
  author={Grimm, Simon L and Stadel, Joachim G},
  journal={The Astrophysical Journal},
  volume={796},
  number={1},
  pages={23},
  year={2014},
  publisher={The American Astronomical Society}
}

@article{rein2012rebound,
  title={REBOUND: an open-source multi-purpose N-body code for collisional dynamics},
  author={Rein, Hanno and Liu, S-F},
  journal={Astronomy \& Astrophysics},
  volume={537},
  pages={A128},
  year={2012},
  publisher={EDP Sciences}
}

@article{tabachnik2000asteroids,
  title={Asteroids in the inner solar system—I. Existence},
  author={Tabachnik, SA and Evans, NW},
  journal={Monthly Notices of the Royal Astronomical Society},
  volume={319},
  number={1},
  pages={63--79},
  year={2000},
  publisher={Blackwell Science Ltd Oxford, UK}
}

@book{box2015time,
  title={Time series analysis: forecasting and control},
  author={Box, George EP and Jenkins, Gwilym M and Reinsel, Gregory C and Ljung, Greta M},
  year={2015},
  publisher={John Wiley \& Sons}
}

@article{alves2025deep,
  title={Deep learning identification of asteroids interacting with gs secular resonances},
  author={Alves, AA and Carruba, V and Delfino, EMDS and Silva, VR and Blasco, L},
  journal={Planetary and Space Science},
  pages={106062},
  year={2025},
  publisher={Elsevier}
}

@article{breiman2001random,
  title={Random forests},
  author={Breiman, Leo},
  journal={Machine learning},
  volume={45},
  pages={5--32},
  year={2001},
  publisher={Springer}
}

@article{smirnov2024comparative,
  title={A comparative analysis of machine learning classifiers in the classification of resonant asteroids},
  author={Smirnov, Evgeny},
  journal={Icarus},
  volume={415},
  pages={116058},
  year={2024},
  publisher={Elsevier}
}

@article{bacu2023assessment,
  title={Assessment of asteroid classification using deep convolutional neural networks},
  author={Bacu, Victor and Nandra, Constantin and Sabou, Adrian and Stefanut, Teodor and Gorgan, Dorian},
  journal={Aerospace},
  volume={10},
  number={9},
  pages={752},
  year={2023},
  publisher={MDPI}
}

@phdthesis{roelofs2019measuring,
	title={Measuring Generalization and overfitting in Machine learning},
	author={Roelofs, Rebecca},
	year={2019},
	school={University of California, Berkeley}
}

@article{brownlee2020identify,
	title={How to identify overfitting machine learning models in scikit-learn},
	author={Brownlee, Jason},
	journal={Machine Learning Mastery},
	volume={27},
	year={2020}
}

@article{liu2021stability,
	title={Stability time-scale prediction for main-belt asteroids using neural networks},
	author={Liu, Chao and Gong, Shengping and Li, Junfeng},
	journal={Monthly Notices of the Royal Astronomical Society},
	volume={502},
	number={4},
	pages={5362--5369},
	year={2021},
	publisher={Oxford University Press}
}

@inproceedings{rosu2021asteroid,
  title={Asteroid image classification using convolutional neural networks},
  author={Rosu, Cosmin and Bacu, Victor},
  booktitle={2021 IEEE 17th International Conference on Intelligent Computer Communication and Processing (ICCP)},
  pages={3--10},
  year={2021},
  organization={IEEE}
}

@article{penttila2021asteroid,
  title={Asteroid spectral taxonomy using neural networks},
  author={Penttil{\"a}, Antti and Hietala, Hilppa and Muinonen, Karri},
  journal={Astronomy \& Astrophysics},
  volume={649},
  pages={A46},
  year={2021},
  publisher={EDP sciences}
}

@article{eckmann1995recurrence,
  title={Recurrence plots of dynamical systems},
  author={Eckmann, Jean-Pierre and Kamphorst, S Oliffson and Ruelle, David and others},
  journal={World Scientific Series on Nonlinear Science Series A},
  volume={16},
  pages={441--446},
  year={1995},
  publisher={World Scientific Publishing}
}

@article{marwan2007recurrence,
  title={Recurrence plots for the analysis of complex systems},
  author={Marwan, Norbert and Romano, M Carmen and Thiel, Marco and Kurths, J{\"u}rgen},
  journal={Physics reports},
  volume={438},
  number={5-6},
  pages={237--329},
  year={2007},
  publisher={Elsevier}
}

@article{gardner1998artificial,
  title={Artificial neural networks (the multilayer perceptron)—a review of applications in the atmospheric sciences},
  author={Gardner, Matt W and Dorling, SR},
  journal={Atmospheric environment},
  volume={32},
  number={14-15},
  pages={2627--2636},
  year={1998},
  publisher={Elsevier}
}

@article{faouzi2020pyts,
  title={pyts: A python package for time series classification},
  author={Faouzi, Johann and Janati, Hicham},
  journal={Journal of Machine Learning Research},
  volume={21},
  number={46},
  pages={1--6},
  year={2020}
}

@inproceedings{srivastava2025classification,
	title={Classification and Comparative Analysis of Hazardous Asteroids Using Machine Learning},
	author={Srivastava, Aarav and Gourisaria, Mahendra Kumar and Saw, Anjishnu and Behara, Dayal Kumar and Jain, Sonal and Sardar, Tanvir Habib},
	booktitle={2025 Third International Conference on Networks, Multimedia and Information Technology (NMITCON)},
	pages={1--6},
	year={2025},
	organization={IEEE}
}

@article{chauhan20182018,
  title={2018 international conference on computing, power and communication technologies (GUCON)},
  author={Chauhan, NK and Singh, K},
  journal={IEEE},
  pages={347--352},
  year={2018}
}

@misc{goodfellow2016deep,
  title={Deep learning},
  author={Goodfellow, Ian},
  year={2016},
  publisher={MIT press}
}

@book{geron2022hands,
	title={Hands-on machine learning with Scikit-Learn, Keras, and TensorFlow},
	author={G{\'e}ron, Aur{\'e}lien},
	year={2022},
	publisher={" O'Reilly Media, Inc."}
}

@article{lecun2015deep,
	title={Deep learning},
	author={LeCun, Yann and Bengio, Yoshua and Hinton, Geoffrey},
	journal={nature},
	volume={521},
	number={7553},
	pages={436--444},
	year={2015},
	publisher={Nature Publishing Group UK London}
}

@article{sze2017efficient,
	title={Efficient processing of deep neural networks: A tutorial and survey},
	author={Sze, Vivienne and Chen, Yu-Hsin and Yang, Tien-Ju and Emer, Joel S},
	journal={Proceedings of the IEEE},
	volume={105},
	number={12},
	pages={2295--2329},
	year={2017},
	publisher={Ieee}
}

@article{wisdom1987chaotic,
	title={Chaotic behavior in the solar system},
	author={Wisdom, Jack},
	journal={Nuclear Physics B-Proceedings Supplements},
	volume={2},
	pages={391--414},
	year={1987},
	publisher={Elsevier}
}

@article{milani1997stable,
	title={Stable chaos in the asteroid belt},
	author={Milani, Andrea and Nobili, Anna M and Kne{\v{z}}evi{\'c}, Zoran},
	journal={Icarus},
	volume={125},
	number={1},
	pages={13--31},
	year={1997},
	publisher={Elsevier}
}

@article{liu2023orbital,
	title={Orbital eccentricity and inclination distribution of main-belt asteroids—The Statistical model revisited},
	author={Liu, Shanhong and Wu, Zhengkai and Yan, Jianguo and Gao, Jian and Huang, Hao and Cao, Jianfeng and Li, Xie and Barriot, Jean-Pierre},
	journal={icarus},
	volume={404},
	pages={115650},
	year={2023},
	publisher={Elsevier}
}

@article{williams1981positions,
	title={The positions of secular resonance surfaces},
	author={Williams, JG and Faulkner, J},
	journal={Icarus},
	volume={46},
	number={3},
	pages={390--399},
	year={1981},
	publisher={Elsevier}
}

@article{migliorini1995interlopers,
	title={Interlopers within asteroid families},
	author={Migliorini, F and Zappal{\`a}, V and Vio, R and Cellino, A},
	journal={Icarus},
	volume={118},
	number={2},
	pages={271--291},
	year={1995},
	publisher={Elsevier}
}

@article{bottke2006yarkovsky,
	title={The Yarkovsky and YORP effects: Implications for asteroid dynamics},
	author={Bottke Jr, William F and Vokrouhlick{\`y}, David and Rubincam, David P and Nesvorn{\`y}, David},
	journal={Annu. Rev. Earth Planet. Sci.},
	volume={34},
	number={1},
	pages={157--191},
	year={2006},
	publisher={Annual Reviews}
}

@article{morbidelli2015dynamical,
	title={The dynamical evolution of the asteroid belt},
	author={Morbidelli, Alessandro and Walsh, Kevin J and O'Brien, David P and Minton, David A and Bottke, William F},
	journal={arXiv preprint arXiv:1501.06204},
	year={2015}
}

@article{wisdom1987urey,
	title={Urey prize lecture: Chaotic dynamics in the solar system},
	author={Wisdom, Jack},
	journal={Icarus},
	volume={72},
	number={2},
	pages={241--275},
	year={1987},
	publisher={Elsevier}
}

@book{quinonero2022dataset,
	title={Dataset shift in machine learning},
	author={Qui{\~n}onero-Candela, Joaquin and Sugiyama, Masashi and Schwaighofer, Anton and Lawrence, Neil D},
	year={2022},
	publisher={Mit Press}
}

@article{amodei2016concrete,
	title={Concrete problems in AI safety},
	author={Amodei, Dario and Olah, Chris and Steinhardt, Jacob and Christiano, Paul and Schulman, John and Man{\'e}, Dan},
	journal={arXiv preprint arXiv:1606.06565},
	year={2016}
}

@article{taori2020measuring,
	title={Measuring robustness to natural distribution shifts in image classification},
	author={Taori, Rohan and Dave, Achal and Shankar, Vaishaal and Carlini, Nicholas and Recht, Benjamin and Schmidt, Ludwig},
	journal={Advances in Neural Information Processing Systems},
	volume={33},
	pages={18583--18599},
	year={2020}
}

@article{dvorak1993survey,
	title={A survey of the dynamics of main-belt asteroids. I},
	author={Dvorak, R and Muller, P and Kallrath, J},
	journal={Astronomy and Astrophysics, vol. 274, 627-641. Article.},
	volume={274},
	year={1993}
}

@inproceedings{davis2006relationship,
  title={The relationship between Precision-Recall and ROC curves},
  author={Davis, Jesse and Goadrich, Mark},
  booktitle={Proceedings of the 23rd international conference on Machine learning},
  pages={233--240},
  year={2006}
}

@article{hanley1982meaning,
  title={The meaning and use of the area under a receiver operating characteristic (ROC) curve.},
  author={Hanley, James A and McNeil, Barbara J},
  journal={Radiology},
  volume={143},
  number={1},
  pages={29--36},
  year={1982}
}

@article{aha1991instance,
  title={Instance-based learning algorithms},
  author={Aha, David W and Kibler, Dennis and Albert, Marc K},
  journal={Machine learning},
  volume={6},
  pages={37--66},
  year={1991},
  publisher={Springer}
}

@inproceedings{dietterich2000ensemble,
  title={Ensemble methods in machine learning},
  author={Dietterich, Thomas G},
  booktitle={International workshop on multiple classifier systems},
  pages={1--15},
  year={2000},
  organization={Springer}
}

@inproceedings{abadi2016tensorflow,
  title={$\{$TensorFlow$\}$: a system for $\{$Large-Scale$\}$ machine learning},
  author={Abadi, Mart{\'\i}n and Barham, Paul and Chen, Jianmin and Chen, Zhifeng and Davis, Andy and Dean, Jeffrey and Devin, Matthieu and Ghemawat, Sanjay and Irving, Geoffrey and Isard, Michael and others},
  booktitle={12th USENIX symposium on operating systems design and implementation (OSDI 16)},
  pages={265--283},
  year={2016}
}

@book{gulli2017deep,
  title={Deep learning with Keras},
  author={Gulli, Antonio and Pal, Sujit},
  year={2017},
  publisher={Packt Publishing Ltd}
}

@book{zhou2012ensemble,
  title={Ensemble methods: foundations and algorithms},
  author={Zhou, Zhi-Hua},
  year={2012},
  publisher={CRC press}
}

@article{schapire2013boosting,
  title={Boosting: Foundations and algorithms},
  author={Schapire, Robert E and Freund, Yoav},
  journal={Kybernetes},
  volume={42},
  number={1},
  pages={164--166},
  year={2013},
  publisher={Emerald Group Publishing Limited}
}

@article{chapman2004hazard,
  title={The hazard of near-Earth asteroid impacts on earth},
  author={Chapman, Clark R},
  journal={Earth and Planetary Science Letters},
  volume={222},
  number={1},
  pages={1--15},
  year={2004},
  publisher={Elsevier}
}

@inproceedings{morbidelli1999origin,
  title={Origin and evolution of near Earth asteroids},
  author={Morbidelli, A},
  booktitle={International Astronomical Union Colloquium},
  volume={172},
  pages={39--50},
  year={1999},
  organization={Cambridge University Press}
}

@article{zolotarev2021dynamic,
  title={On the dynamic evolution of the population of near-Earth asteroids},
  author={Zolotarev, RV and Shustov, BM},
  journal={Astronomy Reports},
  volume={65},
  number={6},
  pages={518--527},
  year={2021},
  publisher={Springer}
}

@article{gladman2000near,
  title={The near-Earth object population},
  author={Gladman, Brett and Michel, Patrick and Froeschl{\'e}, Christiane},
  journal={Icarus},
  volume={146},
  number={1},
  pages={176--189},
  year={2000},
  publisher={Elsevier}
}

@article{granvik2018debiased,
  title={Debiased orbit and absolute-magnitude distributions for near-Earth objects},
  author={Granvik, Mikael and Morbidelli, Alessandro and Jedicke, Robert and Bolin, Bryce and Bottke, William F and Beshore, Edward and Vokrouhlick{\`y}, David and Nesvorn{\`y}, David and Michel, Patrick},
  journal={Icarus},
  volume={312},
  pages={181--207},
  year={2018},
  publisher={Elsevier}
}

@article{duncan1998multiple,
  title={A multiple time step symplectic algorithm for integrating close encounters},
  author={Duncan, Martin J and Levison, Harold F and Lee, Man Hoi},
  journal={The Astronomical Journal},
  volume={116},
  number={4},
  pages={2067--2077},
  year={1998}
}

@article{peterson2009k,
  title={K-nearest neighbor},
  author={Peterson, Leif E},
  journal={Scholarpedia},
  volume={4},
  number={2},
  pages={1883},
  year={2009}
}

@article{geurts2006extremely,
  title={Extremely randomized trees},
  author={Geurts, Pierre and Ernst, Damien and Wehenkel, Louis},
  journal={Machine learning},
  volume={63},
  pages={3--42},
  year={2006},
  publisher={Springer}
}

@inproceedings{everhart1985efficient,
  title={An efficient integrator that uses Gauss-Radau spacings},
  author={Everhart, Edgar},
  booktitle={International Astronomical Union Colloquium},
  volume={83},
  pages={185--202},
  year={1985},
  organization={Cambridge University Press}
}

@article{yoshida1990construction,
  title={Construction of higher order symplectic integrators},
  author={Yoshida, Haruo},
  journal={Physics letters A},
  volume={150},
  number={5-7},
  pages={262--268},
  year={1990},
  publisher={Elsevier}
}

@article{carruba2019machine,
  title={Machine-learning identification of asteroid groups},
  author={Carruba, V and Aljbaae, Safwan and Lucchini, A},
  journal={Monthly Notices of the Royal Astronomical Society},
  volume={488},
  number={1},
  pages={1377--1386},
  year={2019},
  publisher={Oxford University Press}
}

@article{carruba2020machine,
  title={Machine learning classification of new asteroid families members},
  author={Carruba, Valerio and Aljbaae, Safwan and Domingos, RC and Lucchini, A and Furlaneto, P},
  journal={Monthly Notices of the Royal Astronomical Society},
  volume={496},
  number={1},
  pages={540--549},
  year={2020},
  publisher={Oxford University Press}
}

@article{baron2019machine,
  title={Machine learning in astronomy: A practical overview},
  author={Baron, Dalya},
  journal={arXiv preprint arXiv:1904.07248},
  year={2019}
}

@article{carruba2022machine,
  title={Machine learning applied to asteroid dynamics},
  author={Carruba, Valerio and Aljbaae, Safwan and Domingos, RC and Huaman, M and Barletta, W},
  journal={Celestial Mechanics and Dynamical Astronomy},
  volume={134},
  number={4},
  pages={36},
  year={2022},
  publisher={Springer}
}

@article{pandey2019machine,
	title={Machine learning algorithms: a review},
	author={Pandey, Darpan and Niwaria, Kamal and Chourasia, Bharti},
	journal={Mach. Learn},
	volume={6},
	number={2},
	year={2019}
}

@article{lecun1998convolutional,
	title={Convolutional networks for images, speech, and time series},
	author={LeCun, Yann and Bengio, Yoshua},
	journal={The handbook of brain theory and neural networks},
	year={1998}
}

@article{hsueh2019condition,
	title={Condition monitor system for rotation machine by CNN with recurrence plot},
	author={Hsueh, Yumin and Ittangihala, Veeresha Ramesha and Wu, Wei-Bin and Chang, Hong-Chan and Kuo, Cheng-Chien},
	journal={Energies},
	volume={12},
	number={17},
	pages={3221},
	year={2019},
	publisher={MDPI}
}

@article{mathunjwa2022ecg,
	title={ECG recurrence plot-based arrhythmia classification using two-dimensional deep residual CNN features},
	author={Mathunjwa, Bhekumuzi M and Lin, Yin-Tsong and Lin, Chien-Hung and Abbod, Maysam F and Sadrawi, Muammar and Shieh, Jiann-Shing},
	journal={Sensors},
	volume={22},
	number={4},
	pages={1660},
	year={2022},
	publisher={MDPI}
}

@article{janiesch2021machine,
	title={Machine learning and deep learning},
	author={Janiesch, Christian and Zschech, Patrick and Heinrich, Kai},
	journal={Electronic markets},
	volume={31},
	number={3},
	pages={685--695},
	year={2021},
	publisher={Springer}
}

@article{mako2005classification,
  title={Classification of near-earth asteroids with artificial neural network},
  author={Mako, Zoltan and Szenkovits, Ferenc and Garda-Matyas, Edit and Csillik, Iharka},
  journal={Studia Mathematica},
  volume={50},
  number={1},
  pages={85--92},
  year={2005}
}

@inproceedings{bahel2021supervised,
  title={Supervised classification for analysis and detection of potentially hazardous asteroid},
  author={Bahel, Vedant and Bhongade, Pratik and Sharma, Jagrity and Shukla, Samiksha and Gaikwad, Mahendra},
  booktitle={2021 International Conference on Computational Intelligence and Computing Applications (ICCICA)},
  pages={1--4},
  year={2021},
  organization={IEEE}
}

@article{bottke2000understanding,
  title={Understanding the distribution of near-Earth asteroids},
  author={Bottke Jr, William F and Jedicke, Robert and Morbidelli, Alessandro and Petit, Jean-Marc and Gladman, Brett},
  journal={Science},
  volume={288},
  number={5474},
  pages={2190--2194},
  year={2000},
  publisher={American Association for the Advancement of Science}
}

@article{quinlan1986induction,
  title={Induction of decision trees},
  author={Quinlan, J. Ross},
  journal={Machine learning},
  volume={1},
  pages={81--106},
  year={1986},
  publisher={Springer}
}

@article{pedregosa2011scikit,
  title={Scikit-learn: Machine learning in Python},
  author={Pedregosa, Fabian and Varoquaux, Ga{\"e}l and Gramfort, Alexandre and Michel, Vincent and Thirion, Bertrand and Grisel, Olivier and Blondel, Mathieu and Prettenhofer, Peter and Weiss, Ron and Dubourg, Vincent and others},
  journal={the Journal of machine Learning research},
  volume={12},
  pages={2825--2830},
  year={2011},
  publisher={JMLR. org}
}

@book{swamynathan2017mastering,
  title={Mastering machine learning with python in six steps: A practical implementation guide to predictive data analytics using python},
  author={Swamynathan, Manohar},
  year={2017},
  publisher={Springer}
}

@article{patro2015normalization,
  title={Normalization: A preprocessing stage},
  author={Patro, SGOPAL and Sahu, Kishore Kumar},
  journal={arXiv preprint arXiv:1503.06462},
  year={2015}
}

@article{lecun2002gradient,
  title={Gradient-based learning applied to document recognition},
  author={LeCun, Yann and Bottou, L{\'e}on and Bengio, Yoshua and Haffner, Patrick},
  journal={Proceedings of the IEEE},
  volume={86},
  number={11},
  pages={2278--2324},
  year={2002},
  publisher={Ieee}
}

@article{krizhevsky2012imagenet,
  title={Imagenet classification with deep convolutional neural networks},
  author={Krizhevsky, Alex and Sutskever, Ilya and Hinton, Geoffrey E},
  journal={Advances in neural information processing systems},
  volume={25},
  year={2012}
}

@article{chambers1999hybrid,
  title={A hybrid symplectic integrator that permits close encounters between massive bodies},
  author={Chambers, John E},
  journal={Monthly Notices of the Royal Astronomical Society},
  volume={304},
  number={4},
  pages={793--799},
  year={1999},
  publisher={Blackwell Science Ltd Oxford, UK}
}

@article{kingma2014adam,
	title={Adam: A method for stochastic optimization},
	author={Kingma, Diederik P and Ba, Jimmy},
	journal={arXiv preprint arXiv:1412.6980},
	year={2014}
}

@inproceedings{smith2017cyclical,
	title={Cyclical learning rates for training neural networks},
	author={Smith, Leslie N},
	booktitle={2017 IEEE winter conference on applications of computer vision (WACV)},
	pages={464--472},
	year={2017},
	organization={IEEE}
}

@incollection{prechelt2002early,
	title={Early stopping-but when?},
	author={Prechelt, Lutz},
	booktitle={Neural Networks: Tricks of the trade},
	pages={55--69},
	year={2002},
	publisher={Springer}
}

@book{hastie2009elements,
  title={The Elements of Statistical Learning: Data Mining, Inference, and Prediction},
  author={Hastie, Trevor and Tibshirani, Robert and Friedman, Jerome H},
  volume={2},
  year={2009},
  publisher={Springer},
  series={Springer Series in Statistics}
}
\bibliographystyle{aasjournalv7}



\end{document}